\documentclass[a4paper]{article}
\usepackage{epsfig}
\usepackage{amsmath}
\usepackage{amsthm}
\usepackage{ifpdf}
\begin{document}

\theoremstyle{remark}
\newtheorem{remark}{Remark}

\parskip 4pt
\abovedisplayskip 7pt
\belowdisplayskip 7pt

\parindent=12pt

\newcommand{\A}{{\bf A}}
\newcommand{\B}{{\bf B}}
\newcommand{\bco}{{\boldsymbol{:}}}
\newcommand{\blambda}{{\boldsymbol{\lambda}}}
\newcommand{\bmu}{{\boldsymbol{\mu}}}
\newcommand{\bn}{{\bf n}}
\newcommand{\bnabla}{{\boldsymbol{\nabla}}}
\newcommand{\bomega}{{\boldsymbol{\omega}}}
\newcommand{\bsigma}{{\boldsymbol{\sigma}}}
\newcommand{\btheta}{{\boldsymbol{\theta}}}
\newcommand{\bg}{{\bf g}}
\newcommand{\bu}{{\bf u}}
\newcommand{\bU}{{\bf U}}
\newcommand{\bv}{{\bf v}}
\newcommand{\bw}{{\bf w}}
\newcommand{\bzero}{{\bf 0}}
\newcommand{\ct}{{\mathcal{T}}}
\newcommand{\cth}{{\mathcal{T}_h}}
\newcommand{\dsum}{{\displaystyle\sum}}
\newcommand{\D}{{\bf D}}
\newcommand{\e}{{\bf e}}
\newcommand{\F}{{\bf F}}
\newcommand{\G}{{\bf G}}
\newcommand{\g}{{\bf g}}
\newcommand{\Gx}{{{\overrightarrow{\bf Gx}}^{\perp}}}
\newcommand{\I}{{\bf I}}
\newcommand{\intbt}{{\displaystyle{\int_{B(t)}}}}
\newcommand{\intG}{{\displaystyle{\int_{\Gamma}}}}
\newcommand{\into}{{\displaystyle{\int_{\Omega}}}}
\newcommand{\intpb}{{\displaystyle{\int_{\partial B}}}}
\newcommand{\lto}{{L^2(\Omega)}}
\newcommand{\no}{{\noindent}}
\newcommand{\obo}{{\Omega \backslash \overline{B(0)}}}
\newcommand{\obt}{{\Omega \backslash \overline{B(t)}}}
\newcommand{\oo}{{\overline{\Omega}}}
\newcommand{\R}{{\text{I\!R}}}
\newcommand{\T}{{\bf T}}
\newcommand{\V}{{\bf V}}
\newcommand{\w}{{\bf w}}
\newcommand{\x}{{\bf x}}
\newcommand{\Y}{{\bf Y}}
\newcommand{\y}{{\bf y}}

\newpage
\thispagestyle{empty}

\noindent{\Large\bf The motion of a neutrally buoyant particle of
an elliptic shape in two dimensional shear flow: a numerical study}

\bigskip
\normalsize \noindent{Shih-Lin Huang$^a$, Shih-Di Chen$^a$, Tsorng-Whay Pan$^{c,}$\footnote{Corresponding authors. e-mail:  pan@math.uh.edu,  mechang@mail.iam.ntu.edu.tw}, 
Chien-Cheng Chang$^{a,b,1}$, Chin-Chou Chu$^a$ } \vskip 1ex
\noindent{$^a$\small Institute of Applied Mechanics, National Taiwan University, Taipei 106, Taiwan, Republic  of China} \vskip 1ex
\noindent{$^b$\small  Center for Advanced Studies in Theoretical Sciences, National Taiwan University, Taipei 106, Taiwan, Republic  of China}\vskip 1ex
\noindent{$^c$\small Department of Mathematics, University of Houston, Houston, Texas  77204, USA} 

\vskip 2ex
\noindent {\bf Abstract}
\vskip 1ex
In this paper, we investigate the motion of a neutrally buoyant cylinder of an elliptic shape  
freely moving in two dimensional shear flow  by direct numerical simulation. An elliptic shape 
cylinder in shear flow, when initially being placed at the middle between two walls, either keeps 
rotating or has a stationary inclination angle depending on the particle Reynolds number 
$Re=G_r r_a^2/\nu$, where $G_r$ is the shear rate, $r_a$ is the semi-long axis of the elliptic 
cylinder and $\nu$ is the kinetic viscosity of the fluid. The critical particle Reynolds number 
$Re_{cr}$ for the transition from a rotating motion to a stationary orientation depends on the 
aspect ratio $AR=r_b/r_a$ and the confined ratio $K=2r_a/H$ where $r_b$ is the semi-short axis 
of the elliptic cylinder and $H$ is the distance between two walls. Although the increasing of 
either parameters makes an increase in $Re_{cr}$,  the dynamic mechanism is distinct. The $AR$ 
variation causes the change of geometry shape; however, the $K$ variation influences the wall effect.  
The stationary inclination angle of non-rotating slender elliptic cylinder with smaller confined ratio  
seems to depend only on the value of $Re-Re_{cr}$.  An expected equilibrium position of 
the cylinder mass center in shear flow is the centerline between two walls, but when placing 
the particle away from the centerline initially, it migrates either toward an equilibrium 
height away from the middle between two walls or back to the middle depending on the 
confined ratio and particle Reynolds number. 
\vskip 1ex
\noindent{\it keywords:} Shear flow; neutrally buoyant elliptic cylinder; equilibrium height; 
stationary inclination angle; fictitious domain/distributed Lagrange multiplier method.

\section{ Introduction}

The problem of particle motion in shear flow is crucially important in many
engineering fields such as the handling of a fluid-solid mixture in slurry, colloid,
and fluidized bed. Experimentally  Segr\'e and Silberberg \cite{Segre1961, Segre1962}  studied the 
migration of neutrally buoyant spheres in a tube Poiseuille flow  and obtained that 
the particles migrate away from the wall and the centerline to accumulate at about 0.6 
of the tube radius from the central axis.  The experiments of Segr\'e and Silberberg 
\cite{Segre1961, Segre1962}  have had a large influence on fluid mechanics studies of migration 
and lift of particles. Comprehensive reviews of experimental and theoretical works 
have been given by Brenner \cite{Brenner1966}, Cox and Mason \cite{Cox1971}, 
 Feuillebois \cite{Feuillebois1989} and Leal \cite{Leal1980}.
 
Jeffery \cite{Jeffery1922} analyzed the equation of motion of a particle immersed in an unbounded viscous 
fluid under Stokes flow, where the fluid and particle inertia could be completely negligible
relative to viscous forces. According to the equation of motion, he corroborated a periodic 
rotation of an ellipsoid in a simple shear flow.
Concerning the theoretical results of the neutrally buoyant particle migration in linear shear flow,  
Bretherton \cite{Bretherton1962} found an expression for the lift force per unit length on a 
cylinder in an unbounded two-dimensional linear shear flow at small Reynolds number. 
Saffman's lift force \cite{Saffman1965} on a sphere 
of radius $r_a$ in an unbounded linear shear flow with shear rate $G_r$ is 
$F_s=6.46\rho V r_a^2 (G_r\nu)^{1/2}=6.46\rho \nu r_a V (Re)^{1/2}$ where 
$\nu$ is the kinetic viscosity of the fluid, 
$\rho$ is the density of the fluid, $Re=G_r r_a^2/\nu$ is the particle Reynolds number, 
and $V$ is the slip velocity of the sphere.  In a bounded linear shear flow,
Ho and Leal \cite{Ho1974} examined the motion of a rigid sphere  with inclusion of the inertia 
effects at small  Reynolds numbers by a regular perturbation method. The sphere reaches a stable 
lateral equilibrium height which is the midway between the walls. 
Vasseur and Cox \cite{Vasseur1976} also obtained the same stable lateral equilibrium height.
Ho and Leal require that $Re/K^2 \ll 1$ which is more restrictive than the one
$Re/K  \ll 1$ required by  Vasseur and Cox where $K=2 r_a/H$ is the confined ratio,
$H$ being the distance between two walls.
Via direct numerical simulation, Feng {\it et al.} \cite{FengJ1994} investigated the motion
of neutrally buoyant and non-neutrally buoyant circular particle in plane
shear and Poiseuille flows using a finite element method and obtained
qualitative agreement with the results of perturbation theories and of
experiments. The numerical results of a neutrally buoyant circular cylinder 
in a shear flow of $Re=0.625$ have been discussed in details. The cylinder migrates 
back to the midway between two walls due to the wall repulsion at the small Reynolds 
number. They have suggested that  three factors, namely the wall repulsion due to a 
lubrication effect, the slip velocity, and the Magnus type of lift, are possibly responsible
for the lateral migration. 
Ding and Aidun \cite{Ding2000} studied numerically the dynamics of a cylinder of circular 
or elliptic shape suspended in shear flow at various particle Reynolds number.
They obtained the transient from being rotary to stationary as the particle Reynolds 
number is increased for an elliptic cylinder.  Zettner and Yoda \cite{Zettner2001a} followed up the work done by 
Ding and Aidun \cite{Ding2000} and tested the neutrally buoyant  cylinders of elliptic and non-elliptic shape over a 
wide range of aspect ratios within a moderate range of Re and obtained the rotational motion and the stationary 
orientation behavior of a neutrally buoyant rigid body in a simple shear flow qualitatively agree with the results 
of numerical simulations in a substance.

In this paper, we have applied  a distributed Lagrange multiplier/fictitious domain method
developed in \cite{Chen2012, Pan2013} to study the dynamics of a neutrally buoyant elliptic cylinder 
in two-dimensional shear flow. Such methodology has been validated numerically in \cite{Chen2012, Pan2013} 
by comparing with the computational results by Ding and Aidun in \cite{Ding2000} for a  cylinder of circular and 
elliptic shape and the experimental results by  Zettner and Yoda in \cite{Zettner2001} for a circular cylinder. 
The dynamics of a neutrally buoyant particle of elliptic shape has been studied here based on three physical 
parameters: the aspect ratio of the particle $AR$, the degree of confinement $K$, and the particle Reynolds number $Re$. 
Then we investigate on the equilibrium height of a neutrally buoyant elliptic cylinder in shear flow. 
Feng and Michaelides \cite{Feng2003} have investigated the equilibrium heights of non-neutrally and almost neutrally buoyant circular 
cylinders in two-dimensional shear flow. In their simulations, the density ratio between the solid and fluid is 
between 1.005 and 1.1. The equilibrium heights of their lightest circular cylinder (the density ratio of 1.005) 
are far below the centerline for $Re$ between 2 and 4.5. In \cite{Pan2013}, Pan {\it et al.} have obtained 
equilibrium positions off the centerline between two walls, besides the expected one in the middle between 
two walls, for a neutrally buoyant circular cylinder at different Reynolds numbers in shear flow. Such new 
equilibrium height depends on the particle Reynolds number $Re$  and the confined ratio $K$. For a neutrally 
buoyant  cylinder of elliptic shape in shear flow, we have obtained not just similar results when the initial 
position of the mass center is off the middle between two walls. At higher particle Reynolds number, it is not 
surprised to obtain that the neutrally buoyant cylinder of elliptic shape at the off-the-middle equilibrium height 
can keep a stationary inclination angle as those in the middle between two walls since it is still in a shear flow.

The content of this paper is as follows: In Section 2 we briefly introduce a
fictitious domain formulations of the model problem and computational methods associated 
with the neutrally buoyant long particle cases; then in Section
3 we present and discuss the numerical results. The conclusions are summarized in
Section 4.

\section{ A fictitious domain formulation of the model problem}

\begin{figure} [t]
\begin{center}
\leavevmode
\epsfxsize=4.0in
\epsffile{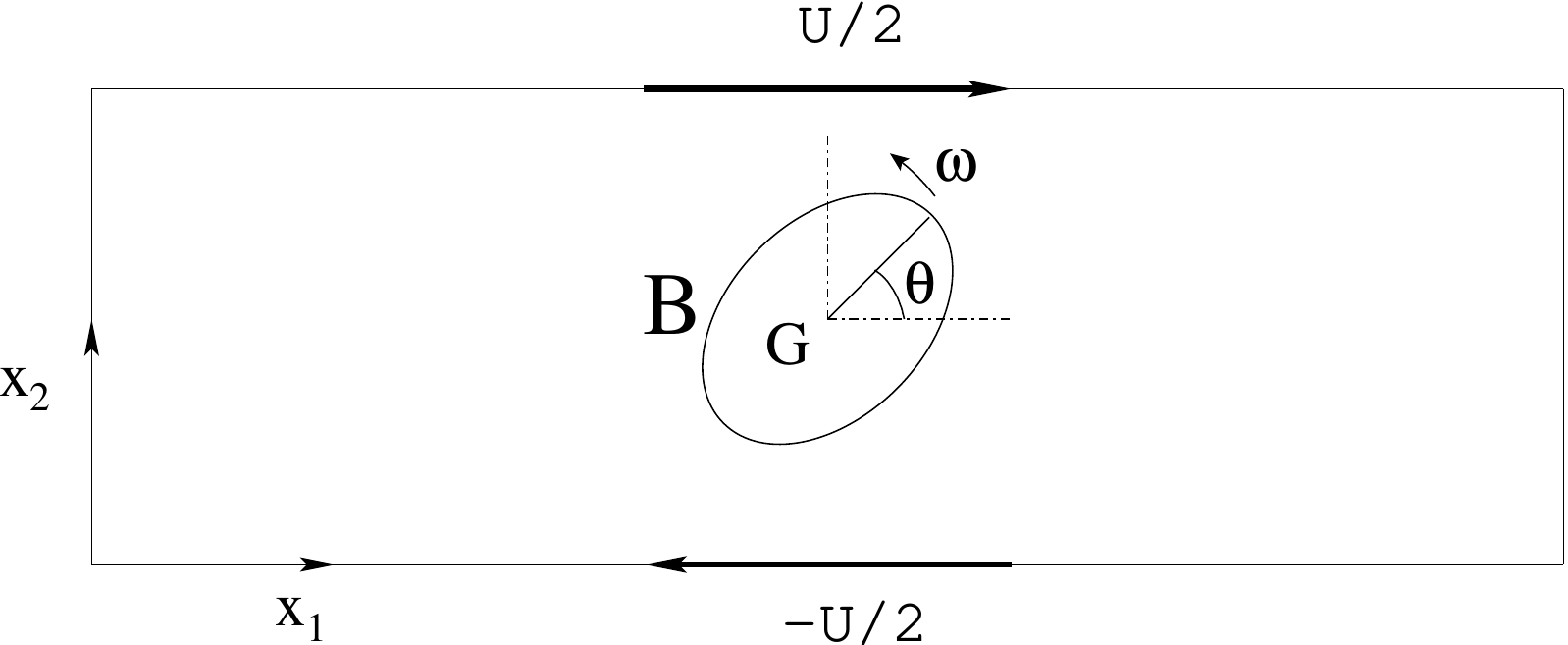}
\end{center}
\caption{ An example of a two-dimensional flow region with a rigid body.}\label{fig:1}
\end{figure}

A fictitious domain formulation with distributed Lagrange multipliers  for flow around freely moving
particles and its associated computational methods have
been developed and tested in, e.g., \cite{RG1999, RG2001, Pan2008, Pan2002, Pan2005, Pan2002a,  Yang2005}. 
For the cases of a neutrally buoyant elliptic cylinder in two dimensional flows, the  
methodology has been developed and validated in \cite{Chen2012, Pan2013}. 
Let $\Omega \subset \R^2$ be a rectangular region
filled with a Newtonian viscous incompressible fluid (of
density $\rho$ and dynamic viscosity $\mu$) and containing
a freely moving neutrally buoyant rigid particle  $B$ centered at $\G=\{G_1, G_2\}^t$
of density $\rho$, as depicted in Figure \ref{fig:1}.  
The flow is modeled by the Navier--Stokes equations and the motion of the particle is modeled 
by the Euler--Newton's equations. The basic idea of
the fictitious domain method is to imagine that the fluid fills the entire space
inside as well as outside the particle boundary. The fluid--flow problem is then posed 
on a larger domain (the ``fictitious domain'').  The fluid inside the particle boundary 
must exhibit a rigid--body motion. This constraint is enforced using the distributed Lagrange 
multiplier, which represents the additional body force per unit volume needed to maintain the 
rigid--body motion inside the particle boundary, much like the pressure in incompressible fluid 
flow, whose gradient is the force required to maintain the constraint of incompressibility. 
For flow around a freely moving neutrally buoyant particle of a long body shape, the fictitious 
domain formulation with distributed Lagrange multipliers is as follows
\begin{eqnarray*}
&&For \ a.e. \ t>0, \ find \
\bu(t) \in W_{{\bf g}_0}, \ p(t) \in L_{0}^2, \ \V_{\G}(t) \in
\R^2, \ \G(t) \in \R^2, \\
&&\omega(t) \in \R, \ \theta(t) \in \R, \ \blambda(t) \in \Lambda_0(t) \ \ \text{\it such that}
\end{eqnarray*}
\begin{eqnarray}
&&\begin{cases}
\rho \into \left[\dfrac{\partial \bu}{\partial t}+(\bu \cdot \bnabla) \bu \right] \cdot \bv\ d\x
+ \mu \into \bnabla \bu : \bnabla \bv\, d\x
- \into p \bnabla \cdot \bv\, d\x \\
\ \ \ = <\blambda, \bv>_{B(t)}, \
\forall \bv \in W_{0},\end{cases} \label{eqn:2.1}\\
&& \into q \bnabla \cdot \bu(t) d\x = 0, \ \forall q \in L^2(\Omega),\label{eqn:2.2}\\
&&<\bmu, \bu(t) >_{B(t)} =0, \ \forall \bmu \in \Lambda_0(t),\label{eqn:2.3}\\
&&\dfrac{d\G}{dt}=\V_{\G},  \label{eqn:2.4} \\
&&\dfrac{d\theta}{dt}=\omega,\label{eqn:2.4a} \\
&& \V_{\G}(0) = \V_{\G}^0, \ \omega(0) = \omega^0, \ \G(0) = \G^0=\{G^0_1, G^0_2\}^t, 
\ \theta(0)=\theta^0,
\label{eqn:2.5}\\
&&\bu(\x, 0) = {\overline \bu}_0(\x) = \begin{cases}
\bu_0(\x), \ \forall \x \in \obo, \\
\V_{\G}^0 + \omega^0 \{-(x_2-G^0_2),x_1-G^0_1\}^t, \ \forall \x \in \overline{B(0)},
\end{cases}
\label{eqn:2.6}
\end{eqnarray}
\noindent where $\bu$ and $p$ denote velocity and pressure, respectively, 
$\blambda$ is a Lagrange multiplier,
$\V_{\G}$ is the translation velocity of the particle $B$, 
$\omega$ is the angular velocity of $B$, 
and $\theta$ is the angle between the horizontal
direction and the long axis of the elliptic cylinder (see Fig. \ref{fig:1}).
We suppose that the no-slip condition holds on $\partial B$.
We also use, if necessary, the notation $\phi(t)$ for the function $\x \to
\phi(\x,t)$. The function spaces in equations (\ref{eqn:2.1})-(\ref{eqn:2.6}) are defined by
\begin{eqnarray*}
&&W_{{\bf g}_0} = \{\bv|\bv \in (H^1(\Omega))^2, \ \bv = {\bf g}_0 \
\text{\it on the top and bottom of $\Omega$ and} \\
&&\hskip 55pt \bv \ \text{\it is periodic in the $x_1$ direction} \},\\
&&W_{0} = \{\bv|\bv \in (H^1(\Omega))^2, \ \bv = {\bf 0} \
\text{\it on the top and bottom of $\Omega$ and} \\
&&\hskip 55pt \bv \ \text{\it is periodic in the $x_1$ direction} \},\\
&&L_{0}^2  = \{q|q \in L^2(\Omega), \int_{\Omega} q\, d\x=0,\},\\
&&\Lambda_0(t) = \{\bmu| \bmu \in (H^1(B(t)))^2, <\bmu,{\bf e}_i>_{B(t)}=0, \ i=1, 2,
<\bmu, \Gx >_{B(t)} = 0\}
\end{eqnarray*}
with ${\bf e}_1=\{1, 0\}^t$, ${\bf e}_2=\{0, 1\}^t$, $\Gx =\{-(x_2-G_2),x_1-G_1\}^t$ and
$<\cdot,\cdot>_{B(t)}$ an inner product on $\Lambda_0(t)$ which can
be the standard inner product on $(H^1(B(t)))^2$.  For simple shear flow, we have
${\bf g}_0=(U/2,0)^t$ on the top wall and $(-U/2,0)^t$ on the bottom wall.

\begin{remark}
The hydrodynamical forces and torque imposed on the rigid body
by the fluid are built in equations (\ref{eqn:2.1})-(\ref{eqn:2.6})
implicitly (see \cite{RG1999, RG2001} for details), thus
we do not need to compute them explicitly in the simulation.
Since in equations (\ref{eqn:2.1})-(\ref{eqn:2.6}) the flow field is
defined on the entire domain $\Omega$, it can be computed
with a simple structured grid.
\end{remark}
\begin{remark}
In equation (\ref{eqn:2.3}), the rigid body motion in the region occupied
by the particle is enforced via the Lagrange multiplier $\blambda$.
To recover the translation velocity $\V_{\G}(t)$  and the angular
velocity $\omega(t)$, we solve the following equations as discussed in
\cite{Pan2002, Pan2013}
\begin{eqnarray}
&&<{\bf e}_i, \bu(t)-\V_{\G}(t)- \omega(t) \ \Gx >_{B(t)}=0,\ for \ i=1, 2,\label{eqn:2.7}\\
&&<\Gx , \bu(t)-\V_{\G}(t)- \omega(t) \ \Gx >_{B(t)}=0.\label{eqn:2.8}
\end{eqnarray}
\end{remark}

\begin{remark}
The method of numerical solution is actually a combination of a distributed Lagrange multiplier based 
fictitious domain method and an operator splitting method. For space discretization, we 
use $P_1$--iso--$P_2$ and $P_1$ finite elements for the velocity
field and pressure, respectively (like in \cite {Bristeau1987}). In time advancing,  
we apply the Lie's scheme in  \cite{Chorin1978} to obtain a sequence of sub--problems 
for each time step (see \cite{Pan2002, Pan2013} for details). 
The computational method has been validated  in \cite{Chen2012, Pan2013} by comparing with the 
computational results by Ding and Aidun in \cite{Ding2000} for a  cylinder of circular and 
elliptic shape and the experimental results by  Zettner and Yoda in \cite{Zettner2001} for a 
circular cylinder. 
\end{remark}

\section{Results and discussions}

\subsection{A neutrally buoyant elliptic cylinder placed initially at the middle between two walls}

\begin{figure}
\begin{center}
\leavevmode
\epsfysize=1.2in
\reflectbox{\epsffile{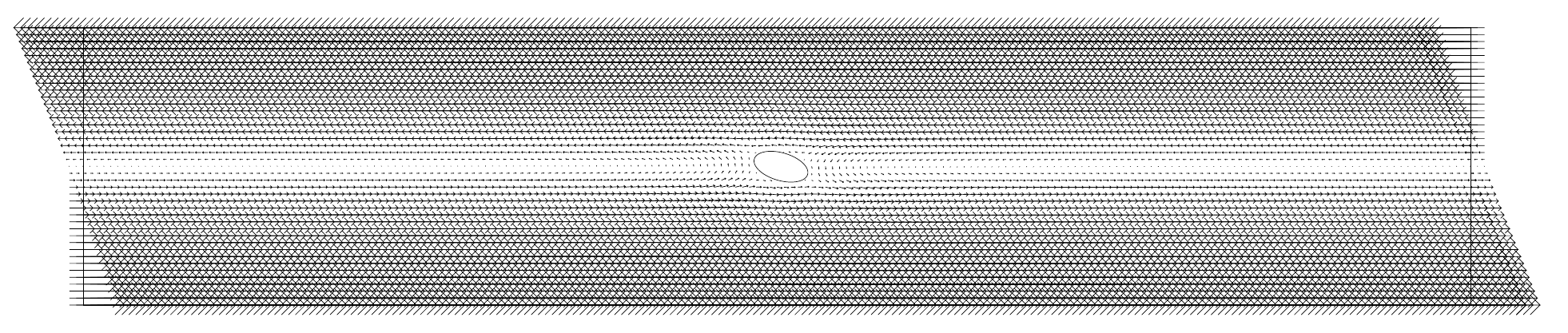}}\\
\epsfysize=1.in
\reflectbox{\epsffile{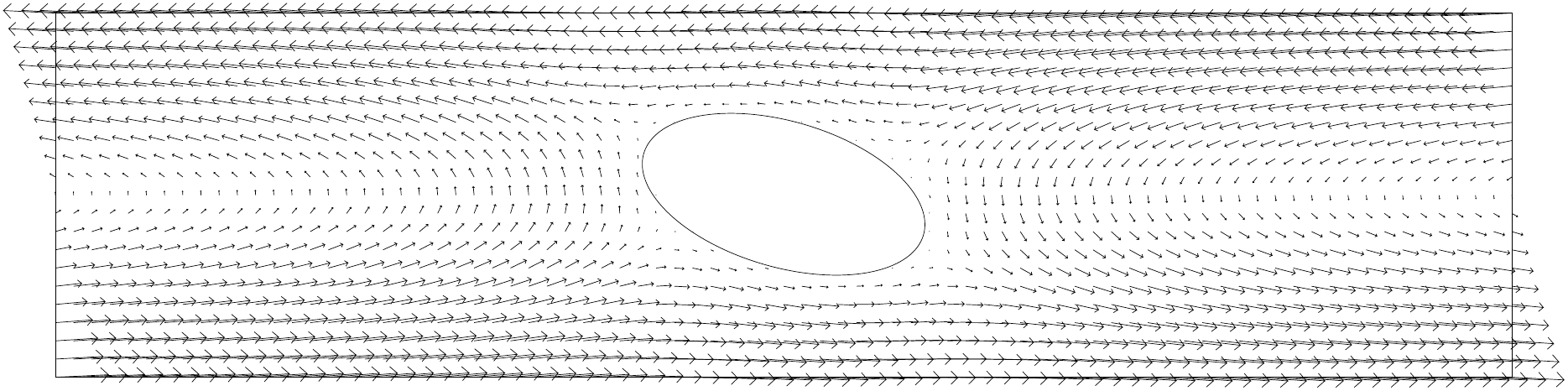}}
\end{center}
\caption{The velocity field (top) and its enlargement (bottom) around an elliptic cylinder 
with a stationary orientation for the case of $K$=0.2 and $Re=8.25$. The critical Reynolds number is $Re_{cr}=7.25$. } \label{fig:2}
\end{figure}

The motion of a neutrally buoyant elliptic cylinder placed initially at the middle between two walls 
in two--dimensional shear flow was studied by Ding 
and Aidun in \cite{Ding2000} via a lattice--Boltzmann method. 
When the particle  Reynolds number is increasing to and less than the critical value, $Re_{cr}$, 
the elliptic shape particle keeps rotating at the middle between two walls but the period of the rotation becomes 
longer and longer. Once $Re$ is beyond the critical value, the elliptic shape cylinder stops rotating but has a 
stationary inclination angle with respect to the wall direction since the torques 
before and after the cylinder are strong enough to hold the particle of long body shape (e.g., see  
Fig. \ref{fig:2} and  \cite{Ding2000}). Such orientation in shear flow with a stationary inclination angle is a
quite different behavior. We have studied the dynamics of such motion based on the particle Reynolds number $Re$, 
the confined ration $K$, and the aspect ratio $AR$ in the following.

\begin{figure}
\begin{center}
\leavevmode
\epsfxsize=2.75in
\epsffile{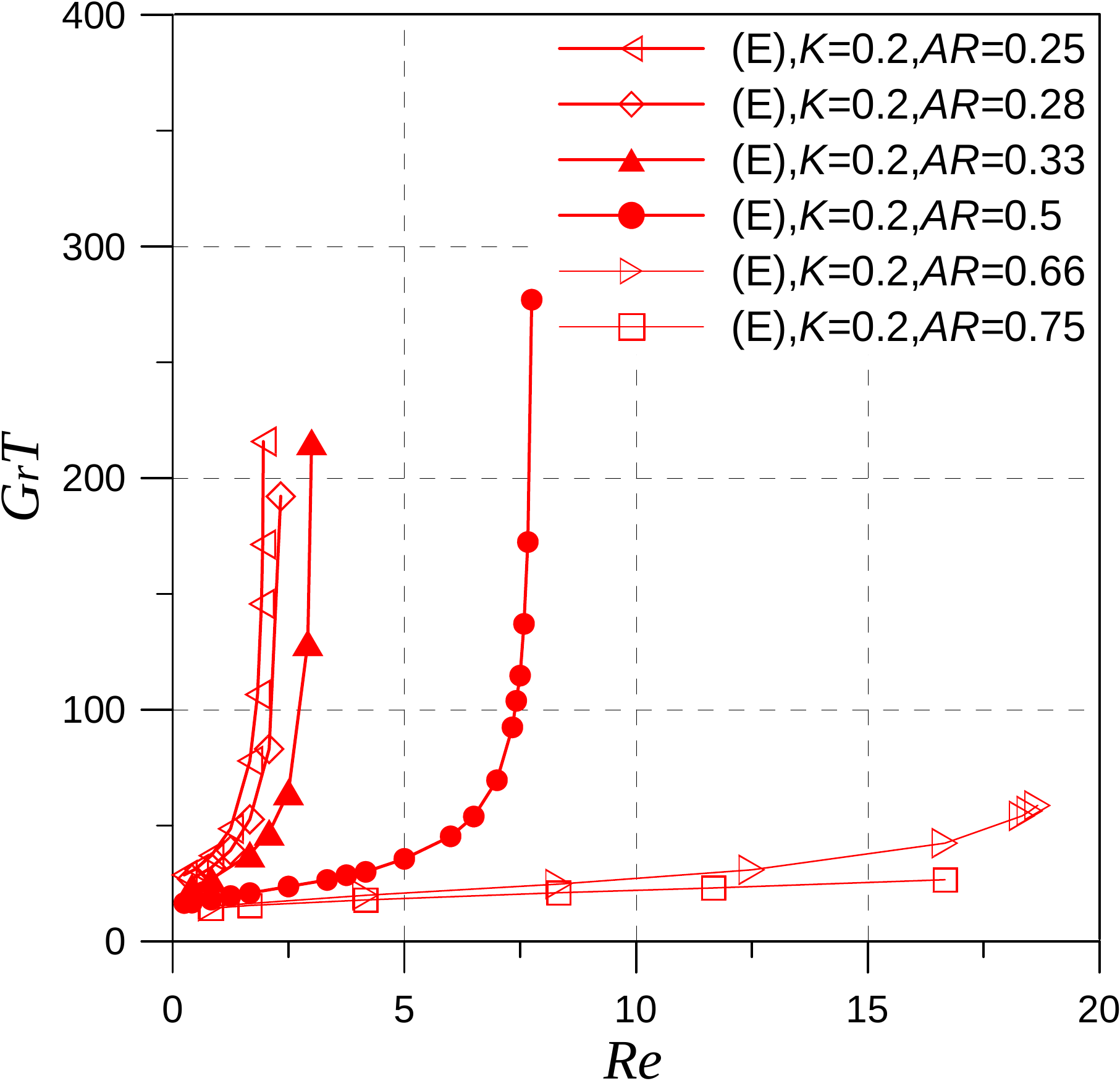}\hskip 5pt
\epsfxsize=2.75in
\epsffile{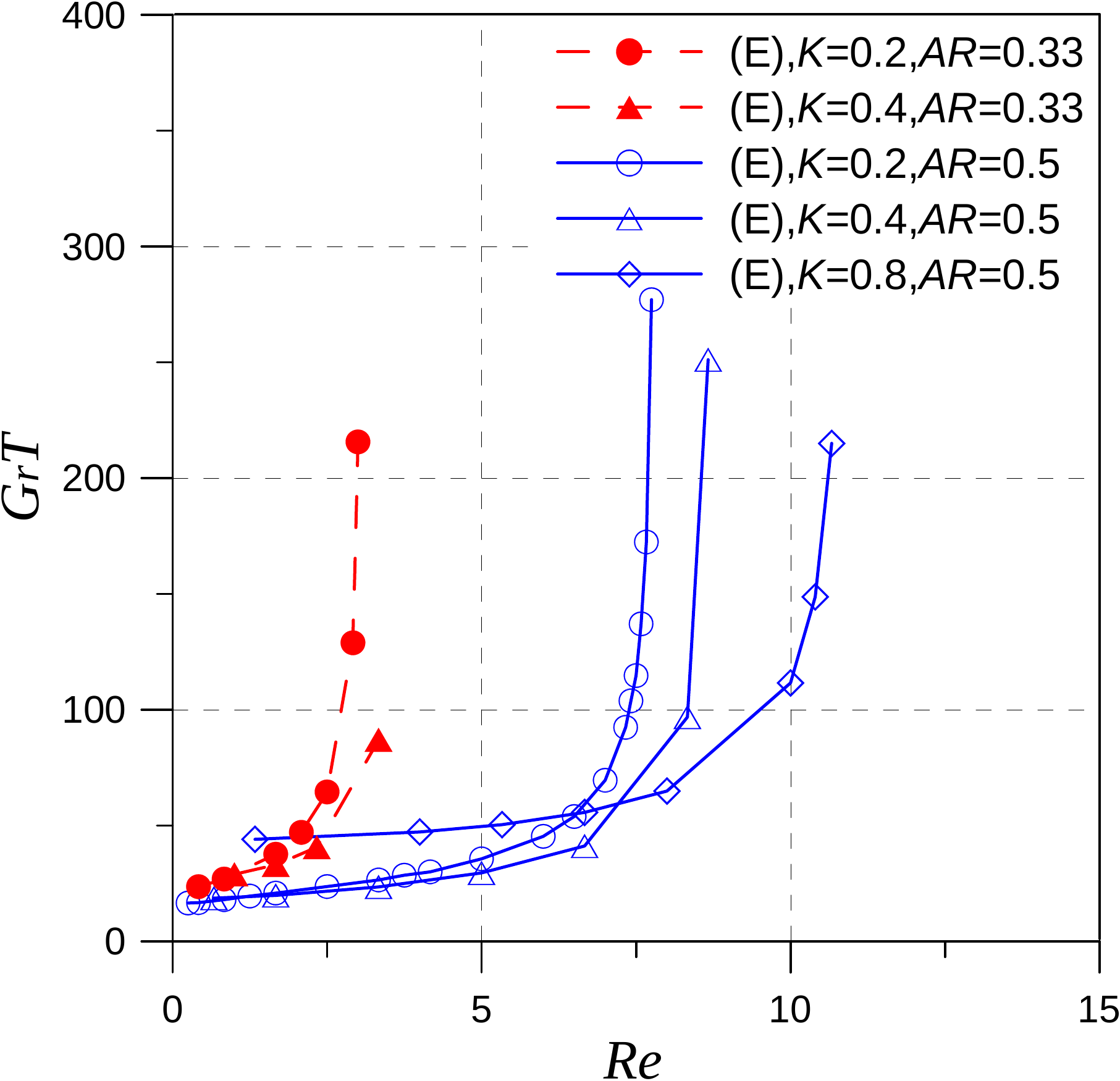}
\end{center}
\caption{The relation between the dimensionless period $G_r T$ and $Re$ of an elliptic cylinder (E) 
with  the variation of the aspect ratio $AR$ under a fixed $K = 0.2$ (left) and the variation of the  confined ratio
$K$ with a fixed aspect ratio $AR$ (right). } \label{fig:3}
\end{figure}

\begin{figure}
\begin{center}
\leavevmode
\epsfxsize=2.75in
\epsffile{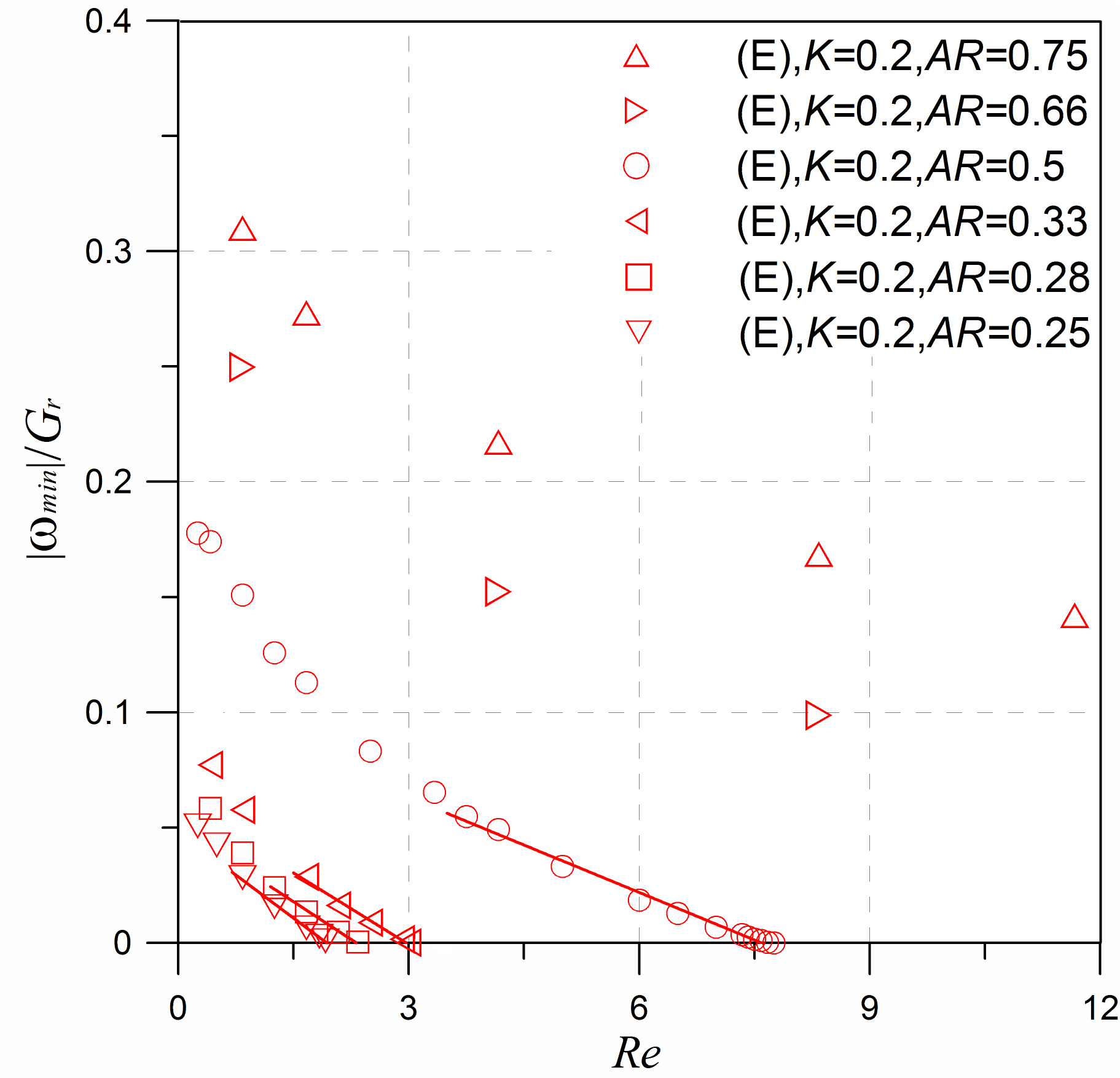} \hskip 5pt
\epsfxsize=2.7in
\epsffile{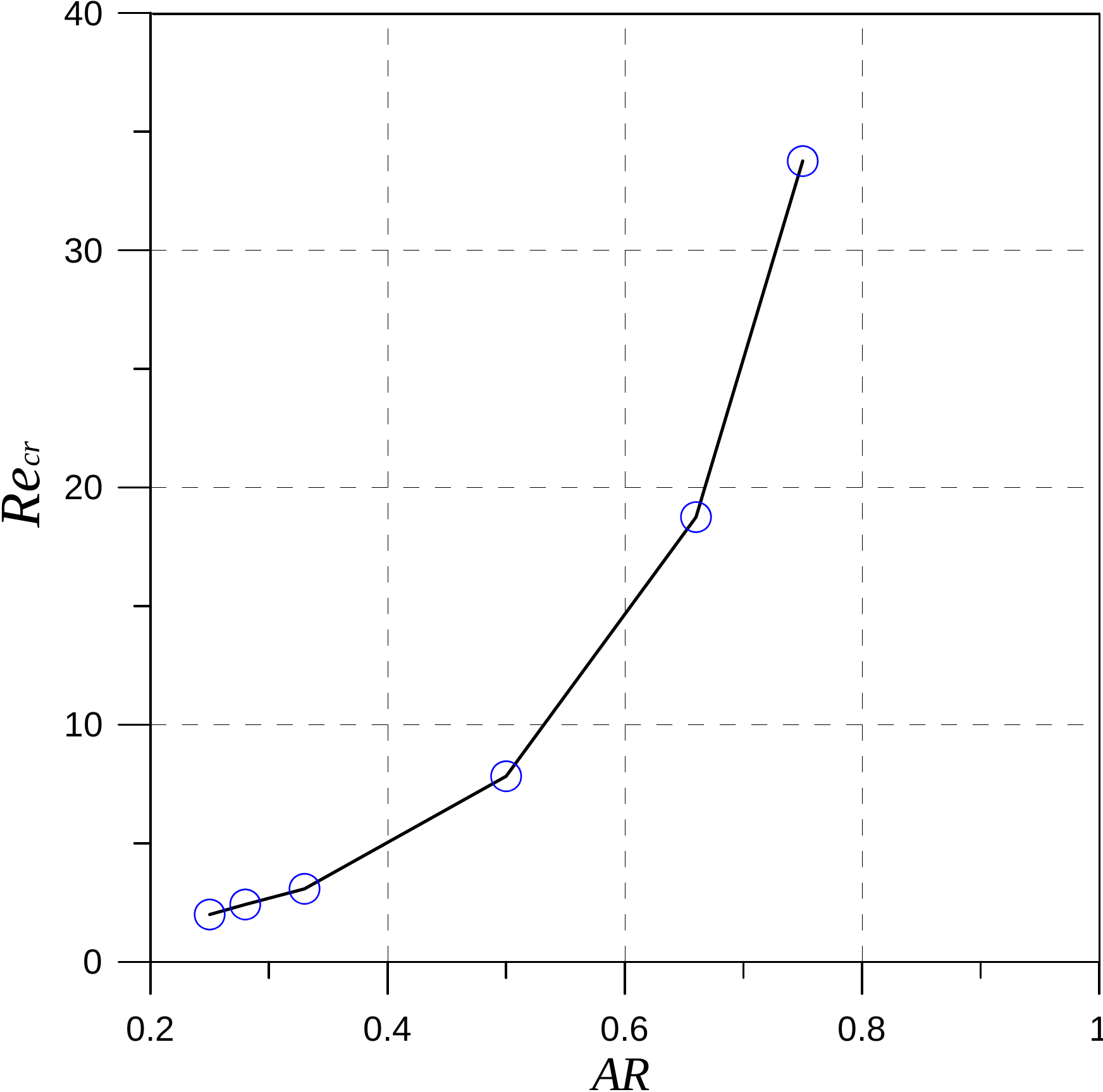} 
\end{center}
\caption{The plot  of the minimum angular velocity $|\omega_{min}/G_r|$ and $Re$ (left) and the plot of the 
critical Reynolds number $Re_{cr}$ versus the aspect ratio $K$ (right) for the cases of an elliptic cylinder (E) with  
the variation of the aspect ratio $AR$ under a fixed $K = 0.2$.} \label{fig:4}
\end{figure} 

In a computational domain $\Omega=[0,5] \times [0,1]$, a neutrally buoyant elliptic cylinder is placed at the 
middle between two walls initially. Thus the distance between two walls is $H=1$. The fluid density is $\rho=1$.
The shear rate is fixed at $G_r=1$
so for a given particle Reynolds number, the kinetic viscosity   is given by $\nu=G_r r_a^2/Re$.
At first, we focus on the rotation of an elliptic cylinder as $0 < Re < Re_{cr}$.
Fig. \ref{fig:3} shows the relation between the dimensionless period $G_r T$ and the Reynolds number $Re$.
The $AR$ effect on the long body with a fixed  $K = 0.2$ in Fig.  \ref{fig:3}  shows that a slender ellipse reveals 
a more rapid increase of $G_r T$ within a small Re than a fat ellipse does within a large $Re$, and its 
$Re_{cr}$ is much smaller than the one associated with a fat ellipse, which is closer to a circular shape.  
In addition, in Fig.  \ref{fig:3}, regarding the $K$ effect under a fixed aspect ratio $AR = 0.33$ and 0.5, 
results show that the less confined long body also present a more rapid increase of $G_rT$ than those of the 
more confined ones and exhibit a smaller $Re_{cr}$. The minimum angular velocity and the Reynolds number 
do show a linear relation  as in Fig.  \ref{fig:4} for the Reynolds number slightly less than the critical value.  
Using such relation, the critical Reynolds number is predicted here and in \cite{Ding2000}. 
Although the increasing of either parameters makes an increase in $Re_{cr}$ (e.g., see Figs.  \ref{fig:3} and  \ref{fig:4}), the dynamic mechanism is distinct. The $AR$ variation causes the change of geometry shape; however, the $K$ 
variation influences the wall effect.

\begin{figure}
\begin{center}
\leavevmode
\epsfxsize=2.9in
\epsffile{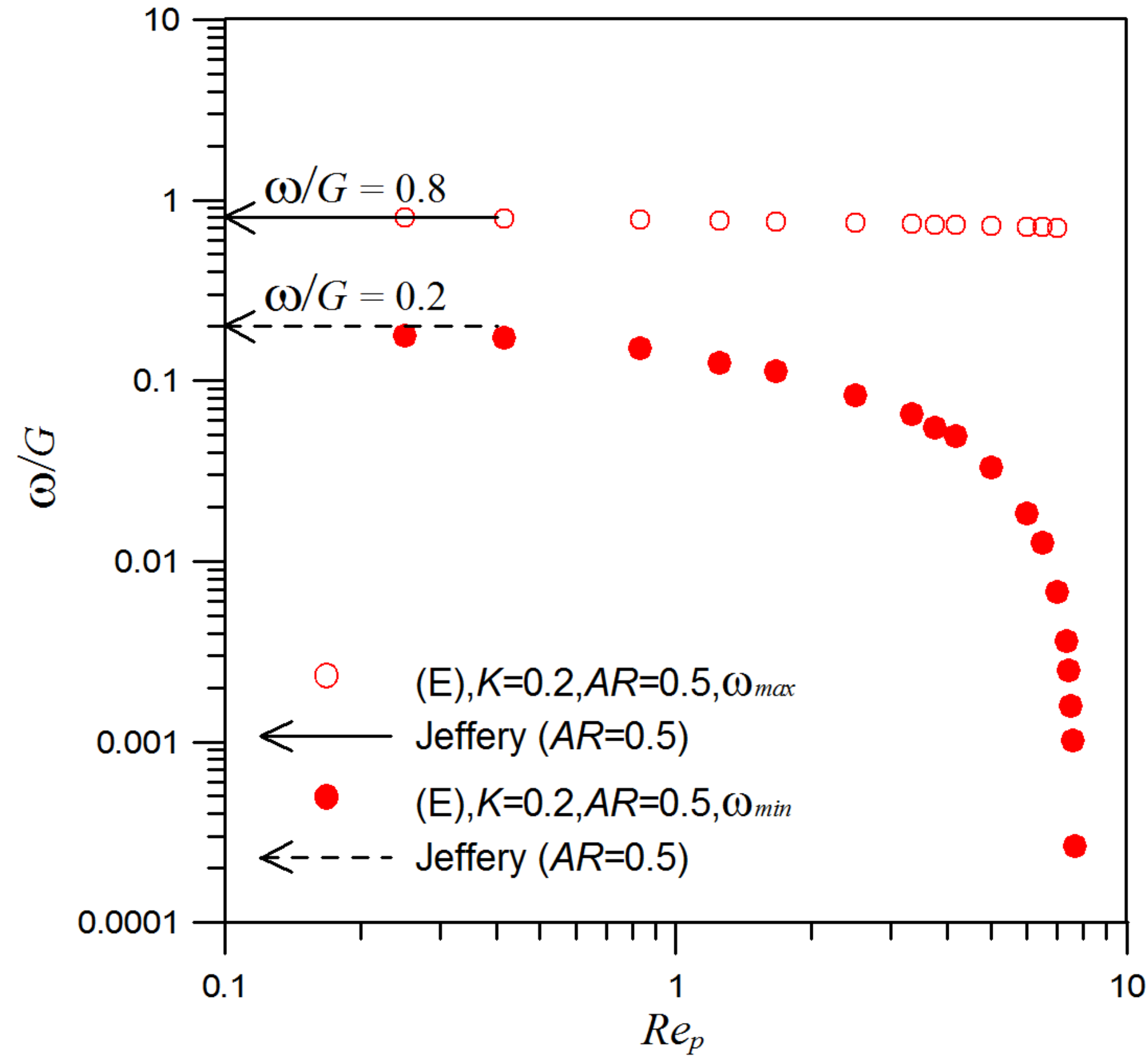}
\end{center}
\caption{Log-log plot of normalized angular velocity $|\omega_{min}|/G_r$ and $|\omega_{max}|/G_r$ versus $Re$
for the fixed aspect ratio $AR=0.5$ and confined ratio $K=0.2$.} \label{fig:5}
\end{figure}

\begin{figure}
\begin{center}
\leavevmode
\epsfxsize=2.75in
\epsffile{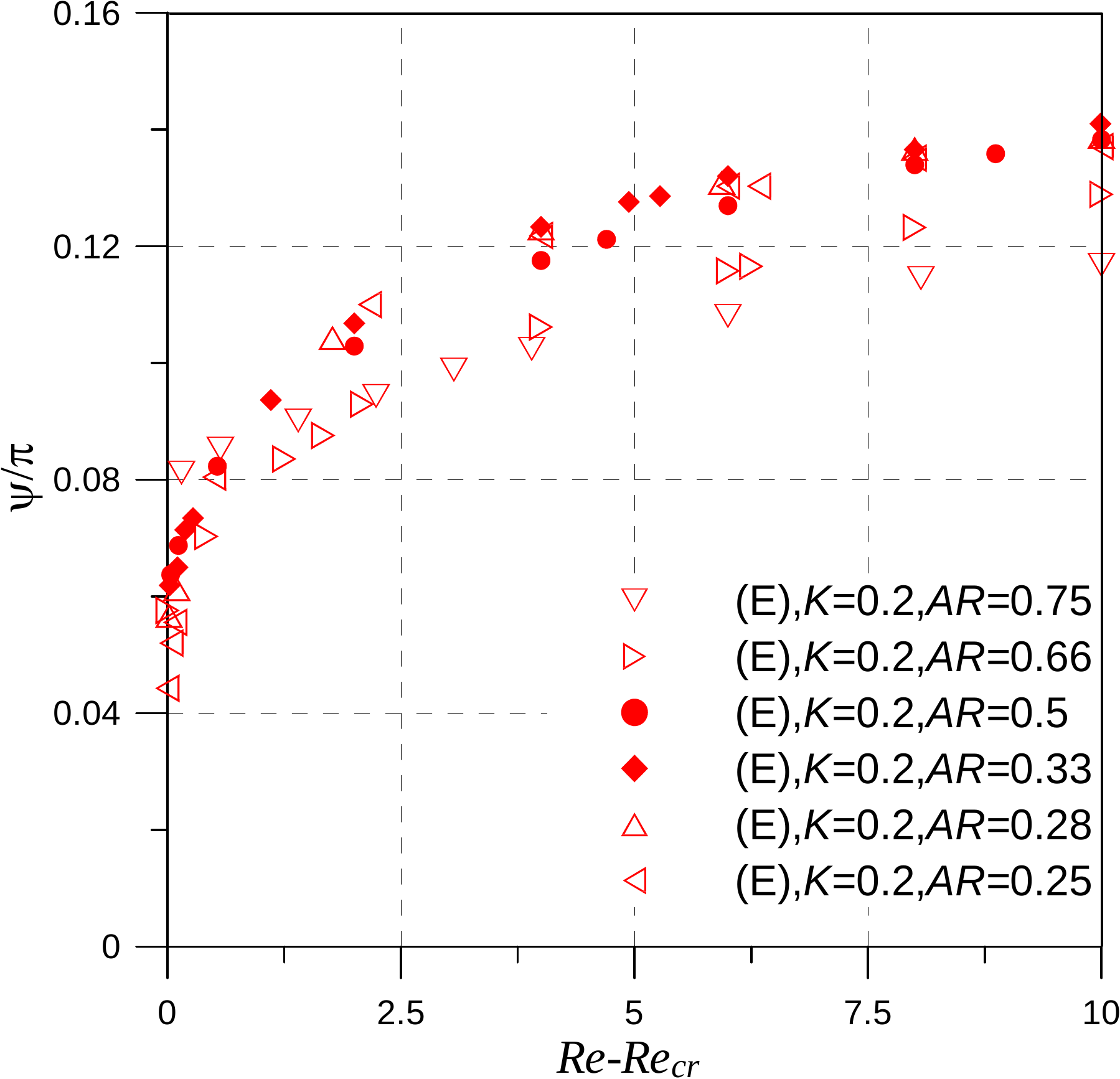} \hskip 5pt
\epsfxsize=2.7in
\epsffile{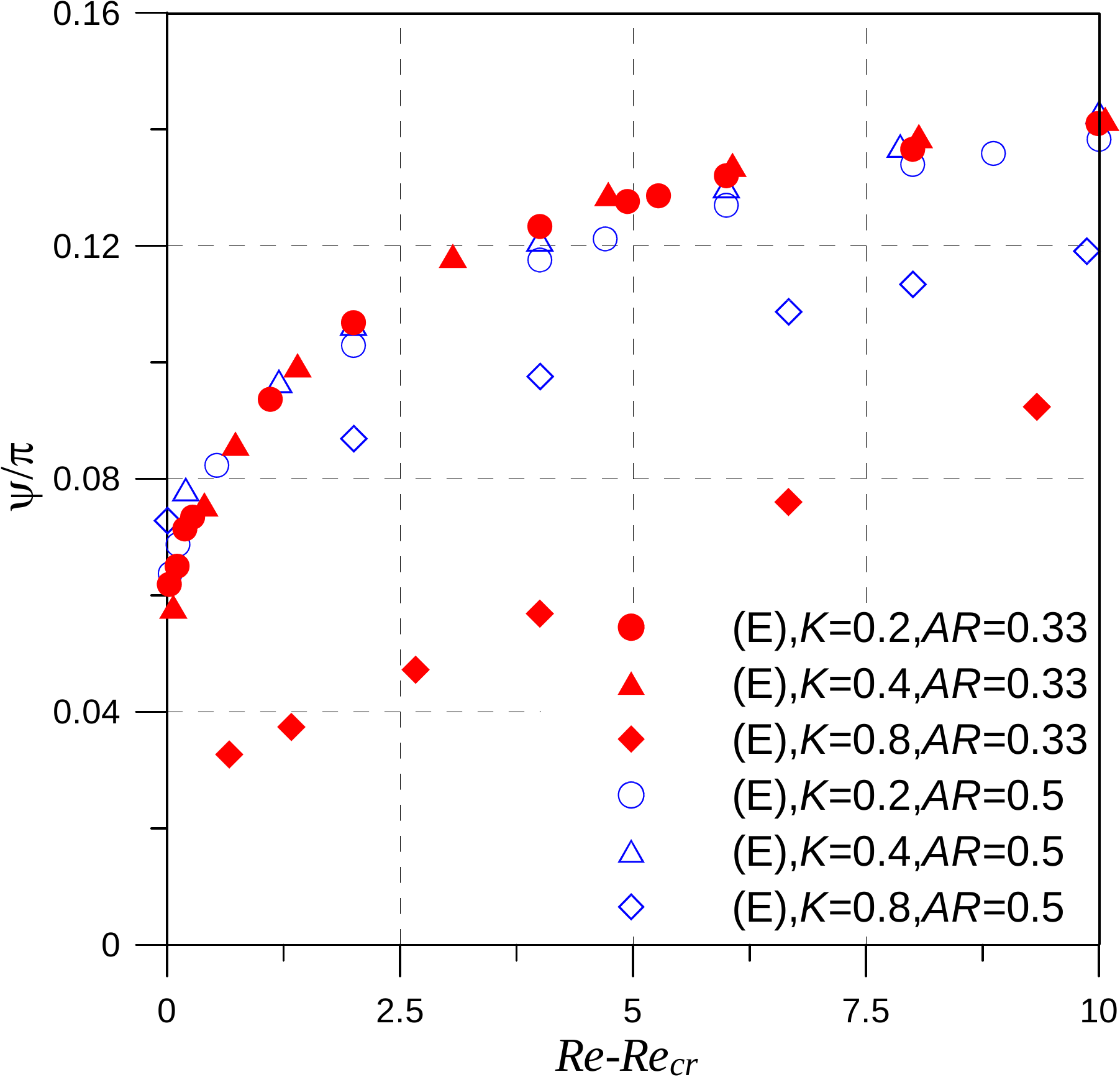} 
\end{center}
\caption{The relation between the inclination angle $\psi$ and the value of $Re-Re_{cr}$ of an elliptic cylinder 
with  the variation of the aspect ratio $AR$ under a fixed $K = 0.2$ (left) and the variation of the  confined ratio
$K$ with a fixed aspect ratio $AR$ (right)} \label{fig:6}
\end{figure} 

Concerning the angular velocity of an elliptic cylinder suspended in shear flow,  
the minimal angular velocity decreases rapidly to about zero when the particle Reynolds 
number $Re$ is closer but less than $Re_{cr}$, e.g., as in   Fig. \ref{fig:5},
since the motion of the elliptic cylinder is about to transit into the one with a 
stationary orientation in  shear flow.   But its maximal angular velocity shown in 
Fig. \ref{fig:5} have different behavior since the maximal angular velocity happens 
when the direction of the long axis is about perpendicular to the shear direction.
For the small particle Reynolds numbers, the maximal and minimal values of the angular 
velocity are very close to the Jeffery's solution as in Fig. \ref{fig:5}.
 
As $Re > Re_{cr}$, an elliptic cylinder in shear flow has a stationary inclination angle. 
We have obtained that the inclination angle  seems to depend only on the value of
$Re-Re_{cr}$ for the aspect ratio less than or equal to 0.5 and the confined ratio less 
than or equal to 0.4 as shown in Fig.  \ref{fig:6}.
But for a fat elliptic cylinder (i.e., $AR> 0.5$) or an elliptic cylinder with higher confined ratio
$K=0.8$, the inclination angle  is smaller. The increase in stationary inclination angle with 
increasing $Re$ have also been obtained by  Ding and Aidun in \cite{Ding2000} and 
Zettner and Yoda in \cite{Zettner2001}. Qualitatively, we have obtained similar behavior.
 
\begin{figure}
\begin{center}
\leavevmode
\epsfxsize=4.in
\epsffile{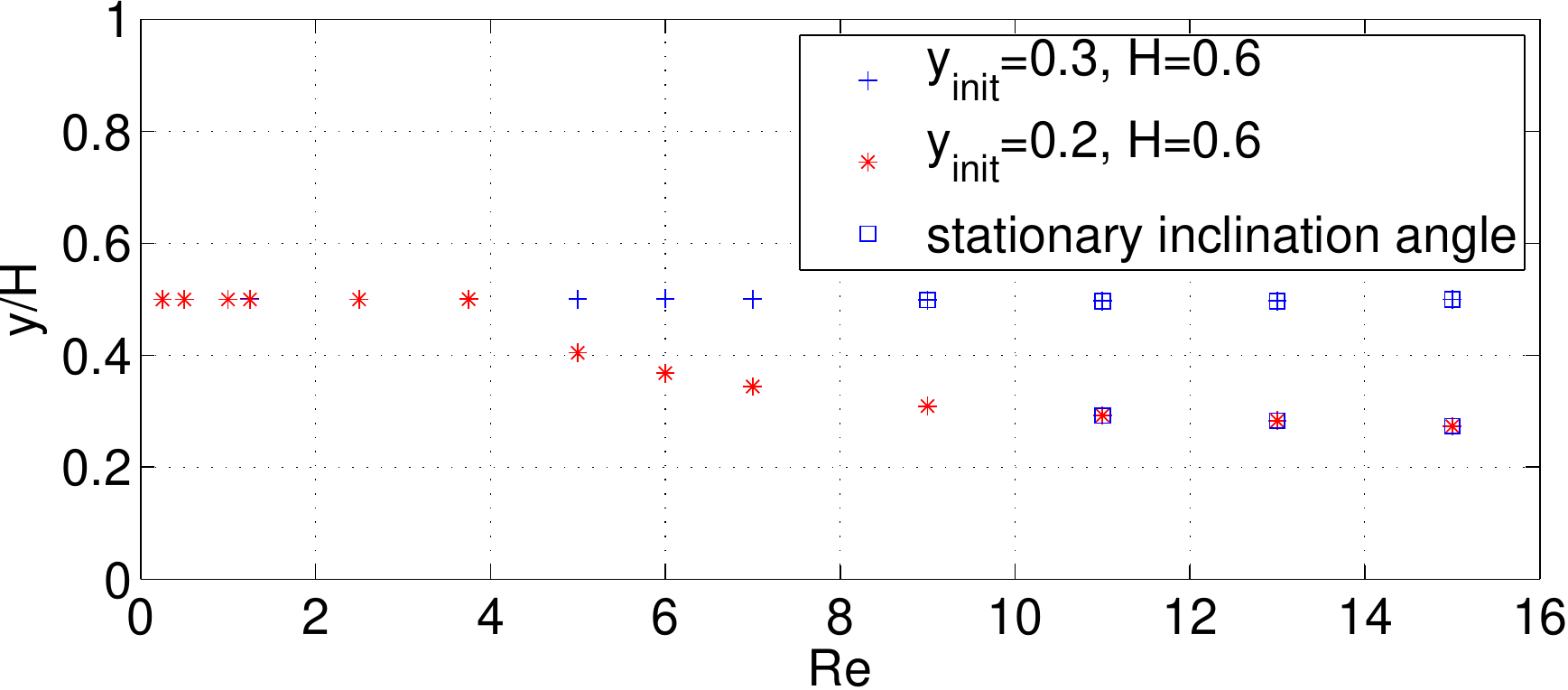} \\
\vskip 1ex
\epsfxsize=4.in
\epsffile{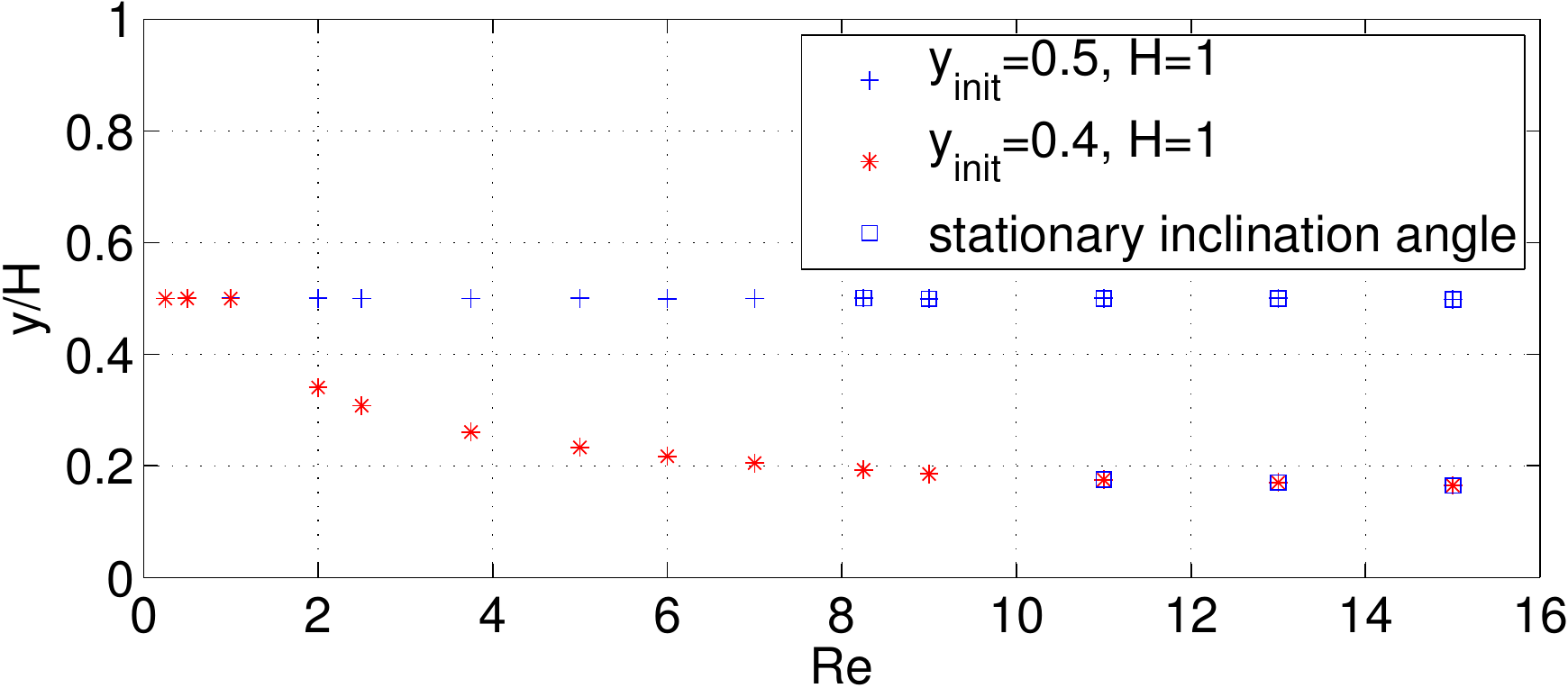} 
\end{center}
\caption{The equilibrium height of the mass center of an elliptic cylinder  versus $Re$ for $H=0.6$ (top)
and 1 (bottom) with $AR=0.5$ and $r_a=0.1$.} \label{fig:7}
\end{figure}

\subsection{A neutrally buoyant elliptic cylinder placed initially away from the midway between two walls}

\begin{figure}[t!]
\begin{center}
\leavevmode
\epsfxsize=2.75in
\epsffile{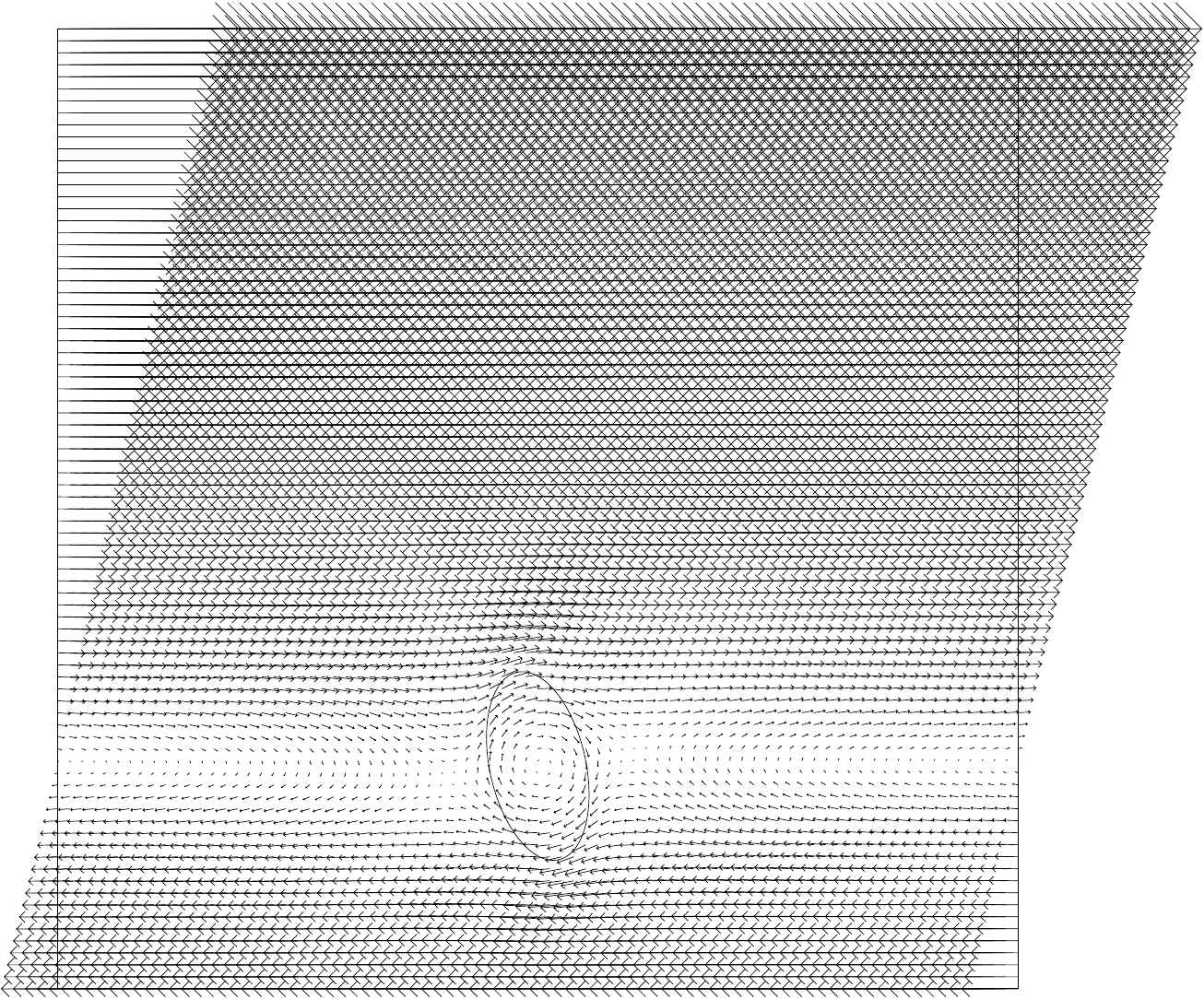} \,
\epsfxsize=2.75in
\epsffile{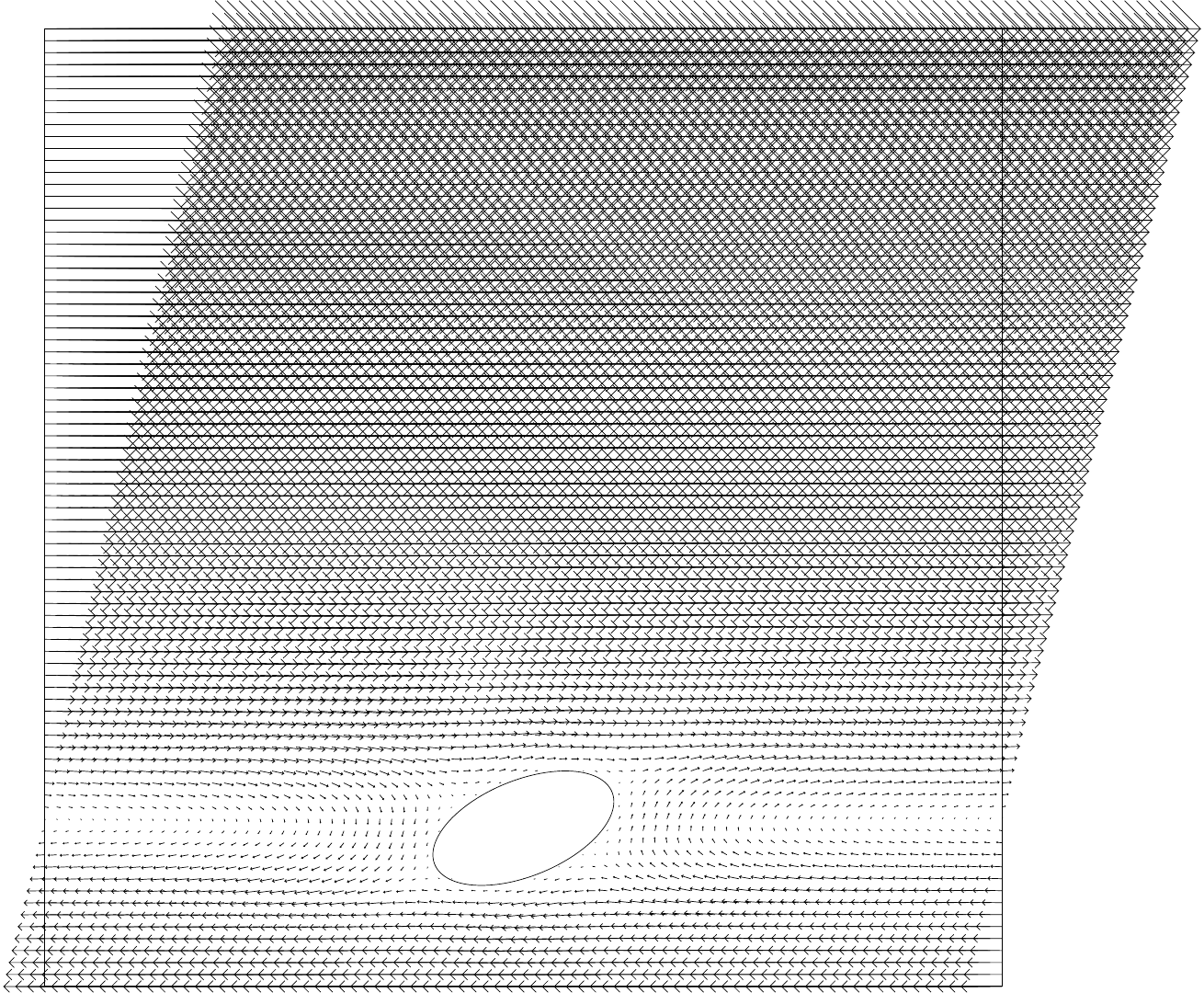} \\
\vskip 4ex
\epsfxsize=2.75in
\epsffile{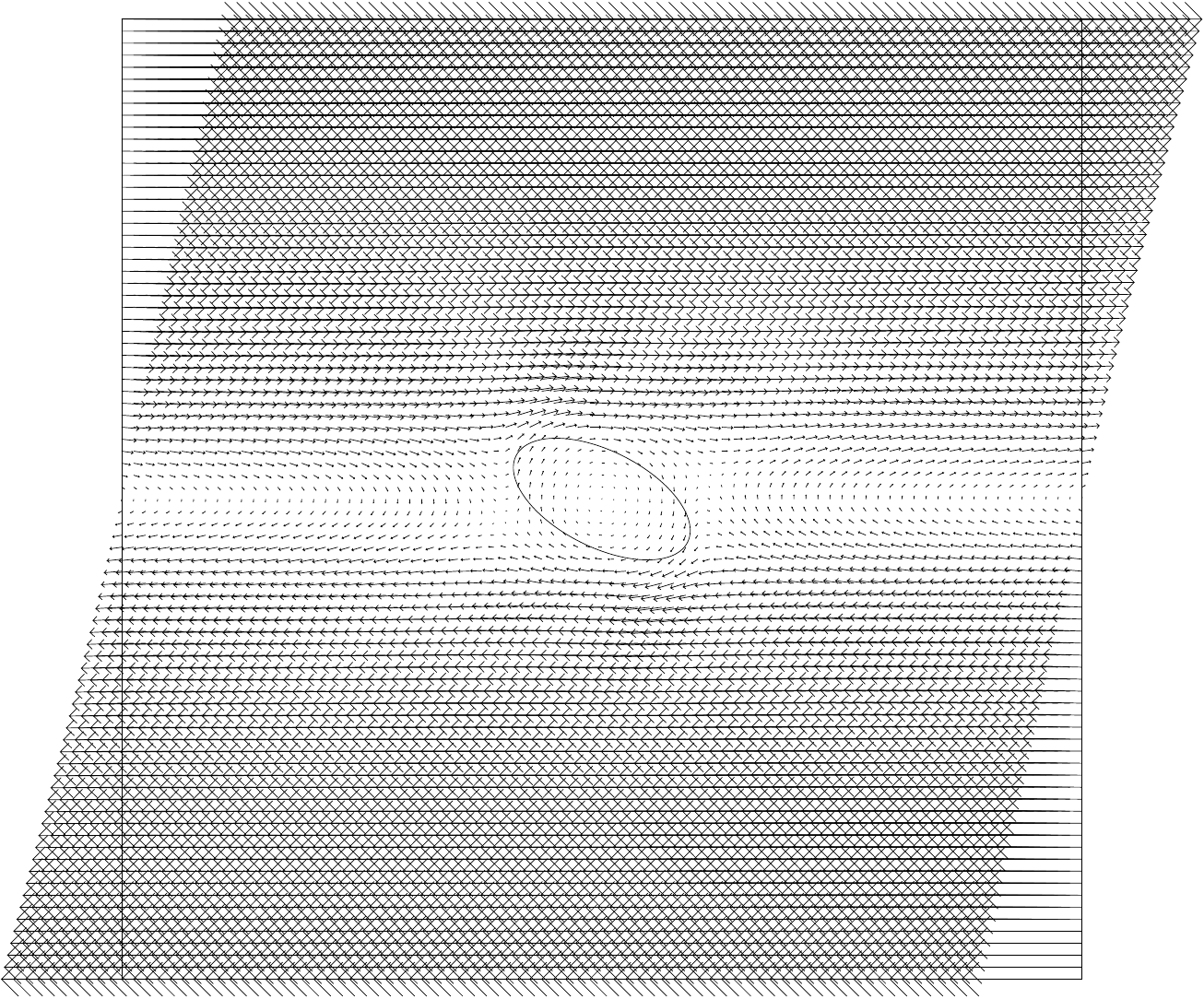} \,
\epsfxsize=2.75in
\epsffile{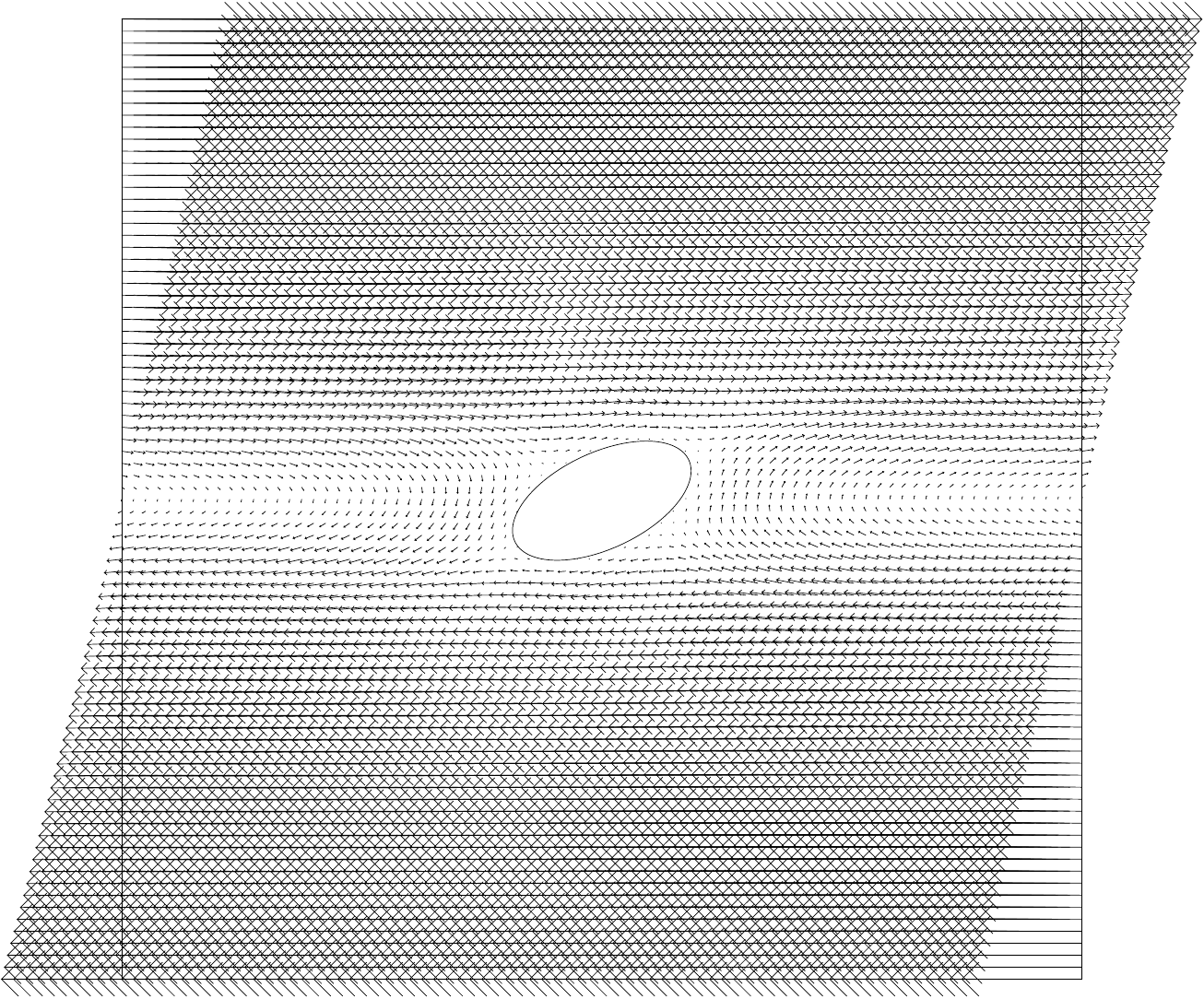} 
\end{center}
\caption{Snapshots of the velocity field around an elliptic cylinder at its equilibrium height obtained by following the 
cylinder mass center for $H=1$ and either the off-the-middle (top) or the midway (bottom) initial position between 
two walls:  at $Re=5$ (two left pictures) the cylinder keeps rotating and at $Re=15$ (two right pictures) the cylinder has a stationary
inclination angle. } \label{fig:8}
\end{figure} 

\begin{figure}[t!]
\begin{center}
\leavevmode
\epsfxsize=4.2in
\epsffile{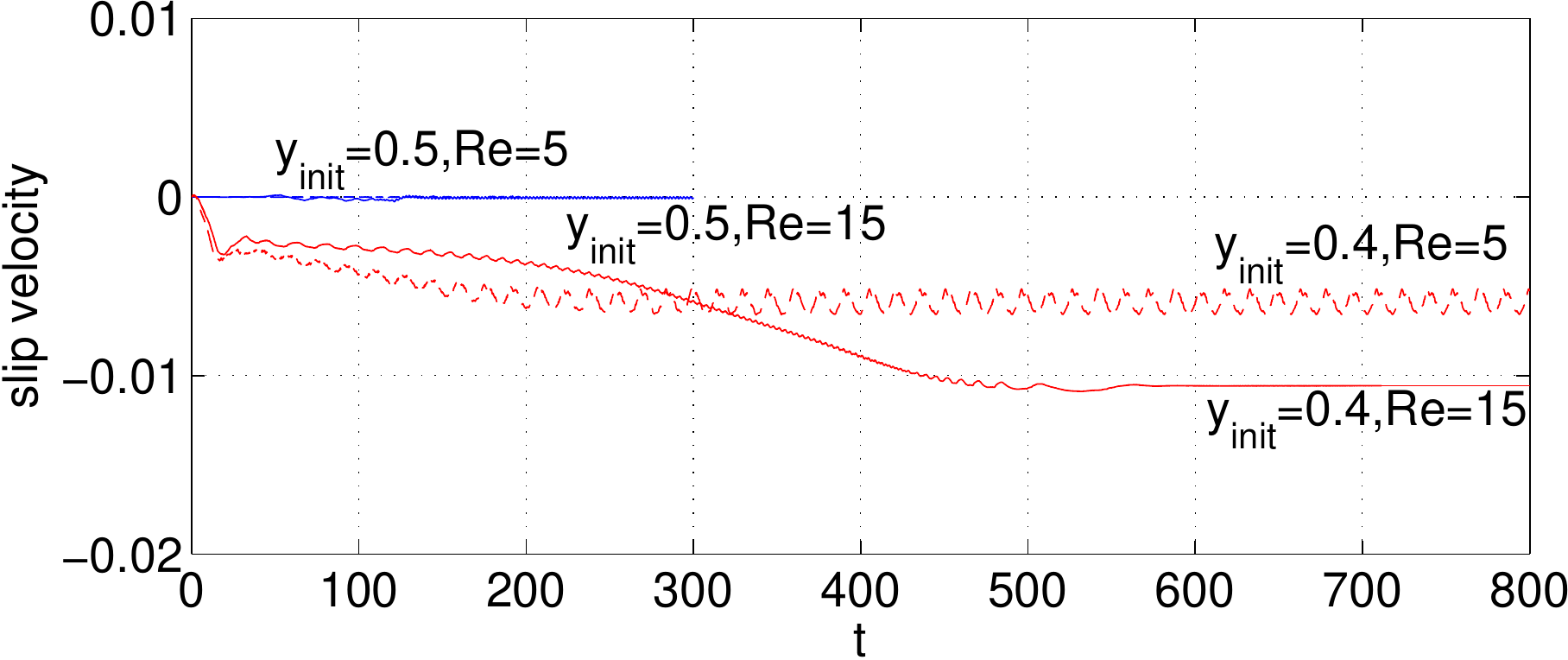}\\
\hskip 15pt \epsfxsize=4.in
\epsffile{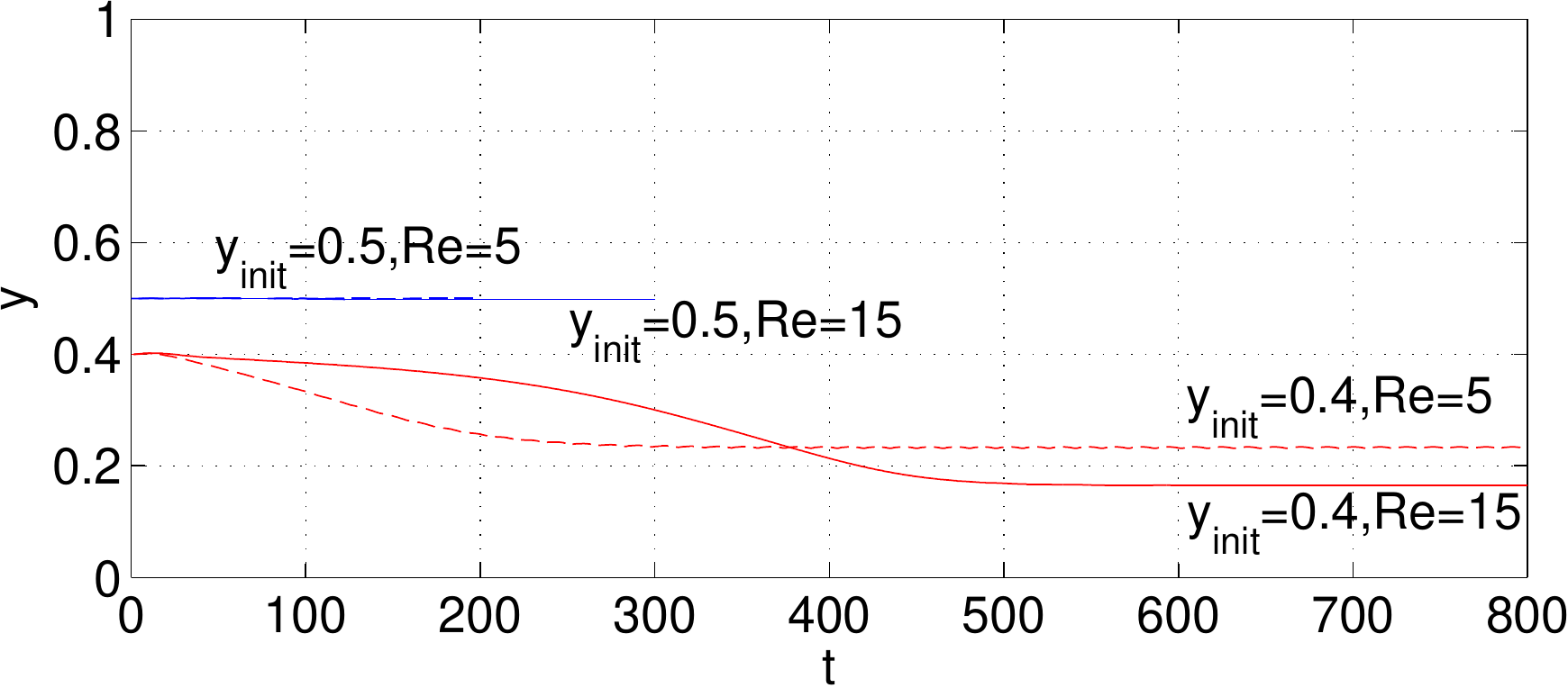}
\end{center}
\caption{Histories of the slip velocity (top) and the height (bottom) of an elliptic cylinder  of  $AR=0.5$ and $r_a=0.1$ with $H=1$ and 
two initial positions: $Re=5$ and $y_{init}=0.5$ (blue dashed-dotted line), $Re=5$ and $y_{init}=0.4$ (red dashed line),
$Re=15$ and $y_{init}=0.5$ (blue solid line), and $Re=15$ and $y_{init}=0.4$ (red solid line)} \label{fig:9}
\begin{center}
\leavevmode
\epsfxsize=4.5in
\epsffile{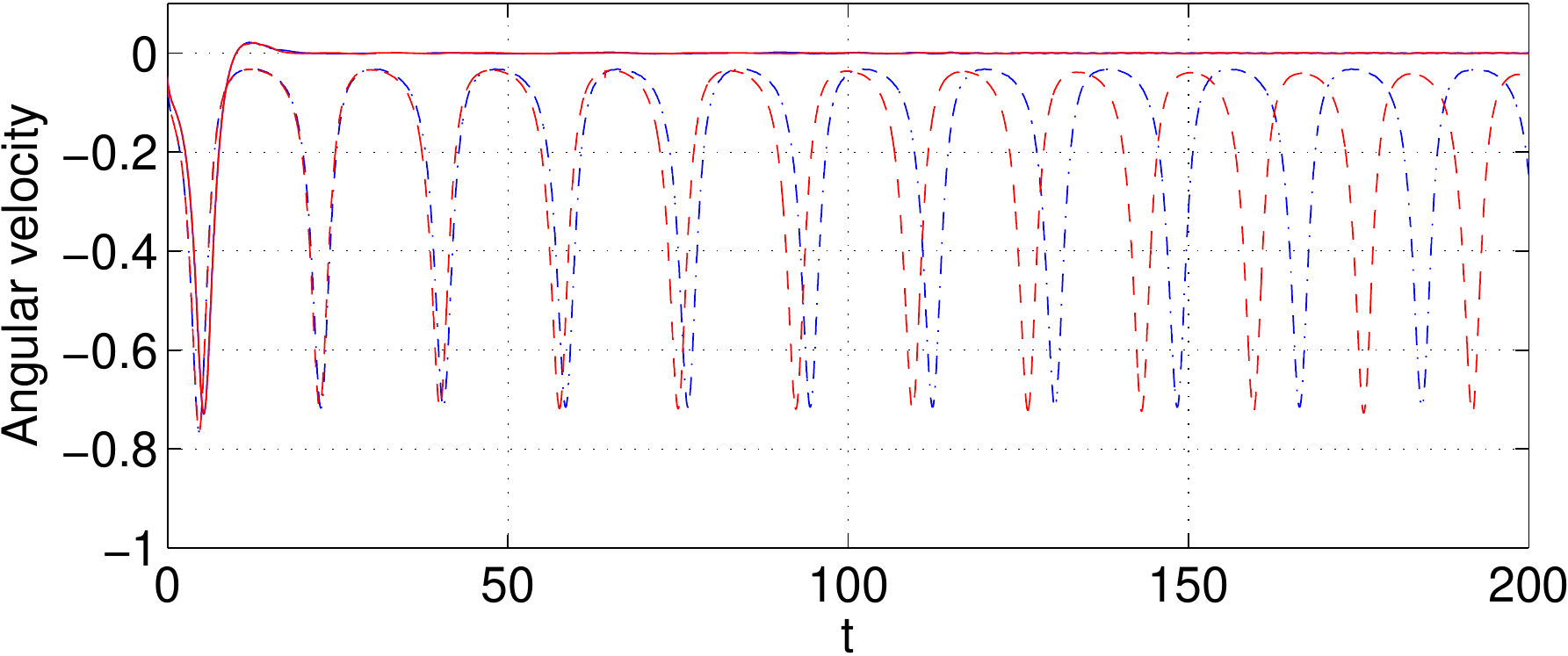}
\end{center}
\caption{Histories of the angular velocity of an elliptic cylinder  of  $AR=0.5$ and $r_a=0.1$ with $H=1$ and 
two initial positions: $Re=5$ and $y_{init}=0.5$ (blue dashed-dotted line), $Re=5$ and $y_{init}=0.4$ (red dashed line),
$Re=15$ and $y_{init}=0.5$ (blue solid line), and $Re=15$ and $y_{init}=0.4$ (red solid line)} \label{fig:10}
\end{figure}

In Ho and Leal \cite{Ho1974} and Vasseur and Cox \cite{Vasseur1976}, they concluded that the 
sphere reaches a stable lateral equilibrium position which is the midway between the walls
for small particle Reynolds numbers. 
In Feng {\it et al.} \cite{FengJ1994}, a circular cylinder migrates back to the midway between two walls 
at $Re=0.625$ when placing it away from the middle between two walls. Feng {\it et al.} have suggested 
that three factors, namely the wall repulsion due to a wall repulsion force, the slip velocity, 
and the Magnus type of lift, are possibly responsible for the lateral migration.
Feng and Michaelides \cite{Feng2003} have investigated the equilibrium heights of non-neutrally buoyant circular 
cylinders in two-dimensional shear flow. In their simulations, the density ratio between the solid and fluid is 
between 1.005 and 1.1. The equilibrium heights of their lighter circular cylinder (the density ratio of 1.005) 
are far below the centerline for $Re$ between 2 and 4.5. In \cite{Pan2013}, 
Pan {\it et al.} have obtained another equilibrium height which is off the middle between two walls for a neutrall 
buoyant circular cylinder in shear flow, besides the one in the middle between two walls.
Such off-the-middle equilibrium height does depend on the particle Reynolds number $Re$  and the confined 
ratio $K$. In the previous section, we have obtained that the midway is always (at least in the range we 
have studied in this paper)  the equilibrium height for the cylinder of elliptic shape when it is 
placed there initially. For the study of the off-the-middle equilibrium height, we have considered a neutrally 
buoyant elliptic cylinder in $\Omega=[0,5]\times[0,H]$  with $r_a=0.1$, $r_b=0.05$, and $H=0.6$ or 1.  
When placing the mass center initially 0.1 length unit below the middle between two walls, the mass center 
of the freely moving cylinder migrates back to the middle between two walls for $Re\le 1$ as in Fig.  \ref{fig:7}; 
but for higher particle Reynolds numbers, the mass center migrates  to  an equilibrium height away from the middle 
as shown in Figs.  \ref{fig:7},   \ref{fig:8} and \ref{fig:9}. It is not surprised to find out that, like 
those staying at the middle, the cylinder either keeps rotating or has a stationary inclination angle when 
being away from the midway since it is still in a shear flow, e.g., see Fig.  \ref{fig:8}. 
For those away from the middle with a stationary
orientation, the inclination angles are 17.62, 20.69 and 22.18 degrees for the particle Reynolds numbers $Re=11$, 13, and 15, 
respectively. Those angles are three to four degrees smaller than the ones associated with the cylinders at 
the middle between two walls at the same Reynolds numbers.  From the snapshots  in Fig. \ref{fig:8}
for $Re=15$, we can see that the smaller inclination angle is due to the effect of the confinement from 
the bottom wall, which is consistent with the results discussed in the previous section.

About the angular velocity, both cylinders keep rotating at $Re=5$ as shown in  Fig. \ref{fig:10} and the period
of the one off the middle initially is slightly smaller due to the effect of the confinement, which is also 
consistent with the results shown in Fig. \ref{fig:3}. At $Re=15$, the histories of the angular velocity of the 
cylinder at two different initial positions are almost identical.  They both rotate right away to their stationary 
inclination angles, respectively, and then stop rotating as indicated in Fig.  \ref{fig:10}. For the one in the 
midway between two walls, it remains there due to the symmetry with respect to the mass center. But for the other 
one not in the middle between two walls initially, Figs.  \ref{fig:9} and \ref{fig:10} show that the cylinder moves 
with a stationary inclination angle  in shear flow  toward the bottom wall slowly and approaches its equilibrium 
height at which, we believe, the force from the effect of slip velocity and the wall repulsive force are balanced as discussed 
in the following.

\begin{figure}
\begin{center}
\leavevmode
\epsfxsize=4.2in
\epsffile{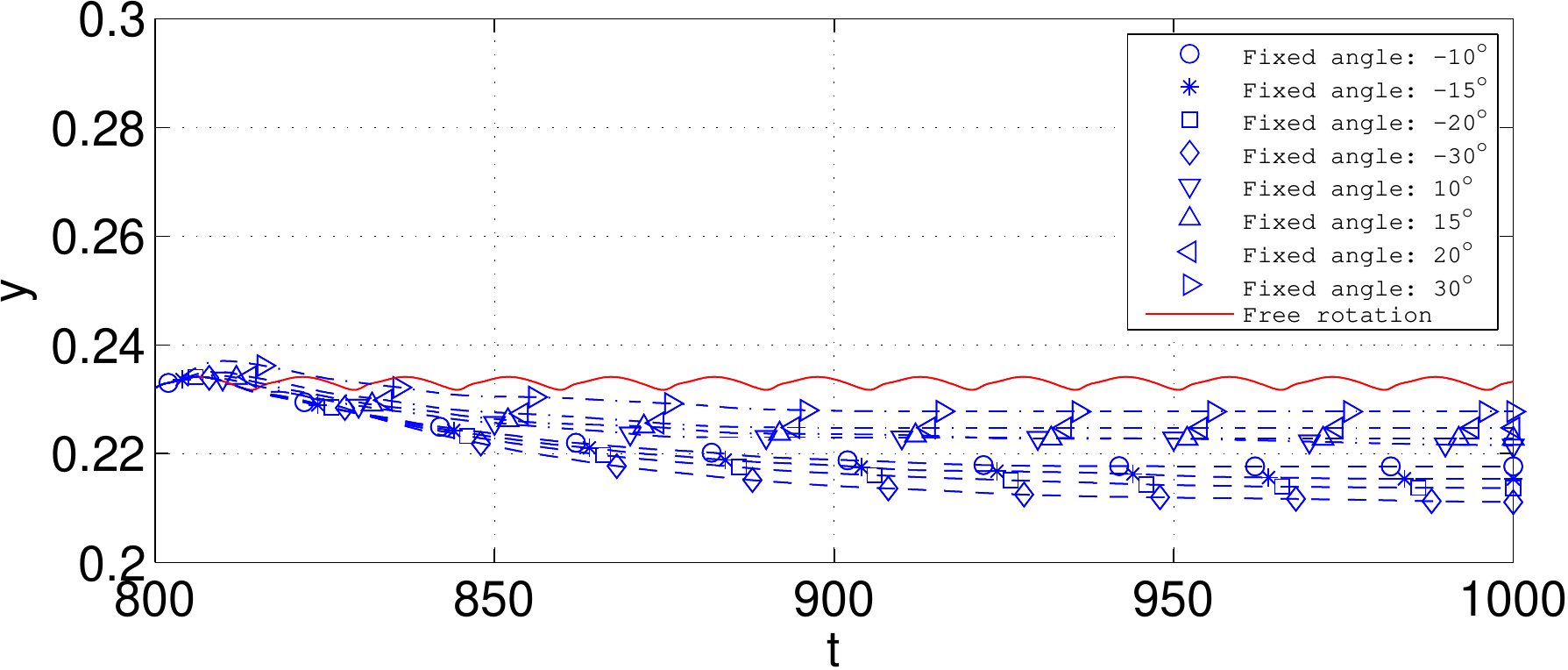}
\end{center}
\caption{Histories of the   the mass center height of an elliptic cylinder with different fixed 
inclination angles, respectively, for $H=1$, $AR=0.5$ and $r_a=0.1$.} \label{fig:11}
\end{figure}

The numerical study on circular cylinders in \cite{Pan2013} about the effect of the Magnus lift force
was done by considering a circular cylinder without rotation in shear flow, which can be achieved by adding
a constraint of zero rotation velocity numerically. Pan {\it et al.} obtained that the mass 
center of a circular cylinder with no rotation is lower than that of the one with totally  free motion,
which is exactly  the effect of the Magnus lift force. In this article, we have studied 
the Magnus lift force by comparing the height of the elliptic cylinder with a fixed orientation angle
to the freely rotating one at $Re=5$. When the inclination angle reaches the specified value 
after $t=800$, the ellipse cylinder is not allowed to rotate any more in shear flow just like the one done in 
\cite{Pan2013}. The chosen inclination angles are $\pm$10, $\pm$15, $\pm$20, and $\pm$30 degrees. The histories 
of the height of the mass center in Fig. \ref{fig:11}   show that the cylinder without rotation goes to a lower 
equilibrium height, which indicates that the Magnus lift force from the rotation does lift a neutrally buoyant 
elliptic shape cylinder in shear flow.

\begin{figure}[t!]
\begin{center}
\leavevmode
\epsfxsize=4.in
\epsffile{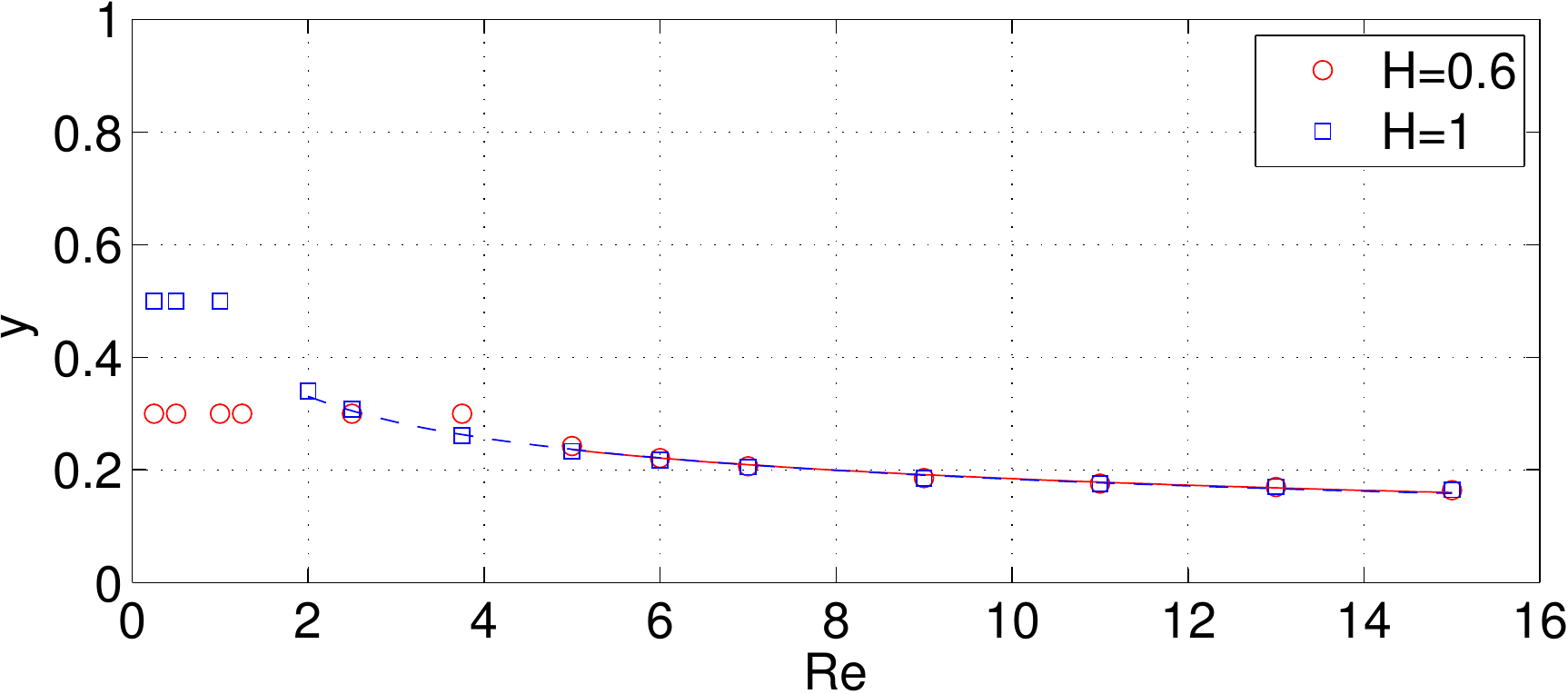}
\end{center}
\caption{The equilibrium height of the mass center of an elliptic cylinder  versus $Re$ for $H=0.6$ and 1 
with $AR=0.5$, $r_a=0.1$, and the initial position away from the middle between two walls. The heights
are functions of the Reynolds number, $y=0.417 Re^{-0.354}$ for $Re\ge 5$ and $y=0.426 Re^{-0.366}$ for $Re\ge 2$,
as $H=0.6$  (red solid line) and 1 (blue dashed line), respectively.} \label{fig:12}
\end{figure} 

Concerning the effect of slip velocity, the cylinder  is moving to the left in the lower region of the 
computational domain in shear flow (see Fig.1 for the set-up of the boundary conditions) due to the initial 
position of the cylinder which is below the centerline. For getting the slip velocity, we first compute the 
fluid horizontal speed on the line through the mass center in front of the cylinder at the distance of 
the half of the computational domain width and then minus the horizontal speed of the cylinder
to obtain the slip velocity. At $Re=5$ and 15, the slip velocities of the cylinder are almost zeros 
for those initially placed at the middle between two walls as in Fig. \ref{fig:9}. But for the ones 
placed initially below the centerline at $Re=5$ and 15, both  slip velocities become negative after 
a very short initial transition period. The negative slip velocity means that the cylinder horizontal
speed to the left is slower than the fluid speed to the left. Thus the cylinder lags the fluid and
then is pushed toward the region with faster flow speed which is the region next to the bottom wall 
for both $Re=5$ and 15. 

Overall the results in Fig. \ref{fig:7}  show that, for $Re \le 1$ (resp., $Re \le 3.75$), the combined effect of the wall repulsion 
force and the Magnus lift force is relatively stronger than the effect of the slip velocity so that the cylinder is pushed back 
to the midway  for $H=1$ (resp., $H=0.6$). But for $Re \ge 2$ (resp., $Re \ge 5$),  the effect of the slip velocity 
dominates the other two so that the cylinder stays away from the middle as $H=1$ (resp., $H=0.6$). 
For those having stationary inclination angle, the balance between the effect of the slip velocity and 
the wall repulsion force  determines the off-the-middle equilibrium height of the cylinder in shear 
flow when the initial position is not at the middle of two walls.
For the actual distance from the cylinder mass center to the bottom wall, Fig. \ref{fig:12} shows that the 
off-the-middle equilibrium heights are about the same for $Re \ge 6$ for both values of $H$. The off-the-middle 
height is a function of the Reynolds number,  $y=0.426 Re^{-0.366}$ for $Re\ge 2$ 
(resp., $y=0.417 Re^{-0.354}$ for $Re\ge 5$) as $H=1$ (resp., $H=0.6$).


\section{Conclusions}

We have investigated the motion of a neutrally buoyant cylinder of an elliptic shape in two dimensional 
shear flow of a Newtonian fluid by direct numerical simulation.  An elliptic shape cylinder in shear 
flow, when initially being placed at the middle between two walls, either keeps rotating or has a 
stationary inclination angle depending on the particle Reynolds number $Re$. The critical particle 
Reynolds number $Re_{cr}$ for the transition from a rotating motion to a stationary orientation 
depends on the aspect ratio $AR$ and the confined ratio $K$. Although the increasing of 
either parameters makes an increase in $Re_{cr}$,  the dynamic mechanism is distinct. The $AR$ 
variation causes the change of geometry shape; however, the $K$ variation influences the wall effect.  
The stationary inclination angle of non-rotating slender elliptic cylinder with smaller confined ratio  
seems to depend only on the value of $Re-Re_{cr}$.  An expected equilibrium position of 
the cylinder mass center in shear flow is the centerline between two walls; but when placing 
the particle away from the centerline initially, it either migrates 
back to the midway or moves away from the middle between two walls depending on the 
particle Reynolds number and the confined ratio.  For those having stationary inclination angle, 
the balance between the effect of the 
slip velocity and the wall repulsion force does play a role for determining the equilibrium height of 
the cylinder in shear flow when the initial position is not at the middle  of two walls.

\vskip 2ex
\noindent{\large\bf Acknowledgments.} T.-W. Pan acknowledges the support 
by the US NSF under Grant No. DMS-0914788. 
S.-L. Huang, S.-D. Chen, C.-C. Chu, C.-C. Chang acknowledge the support by 
the National Science Council (Taiwan, ROC) under Contract Numbers, 
NSC97-2221-E-002-223-MY3, NSC99-2628-M-002-003 and NSC100-2221-E-002-152-MY3.

\end{document}